\documentclass[prd,superscriptaddress,amsfonts,amssymb,amsmath,showpacs,twocolumn]{revtex4-2}

\usepackage{bm}
\usepackage{amsfonts}
\usepackage{latexsym}
\usepackage[latin1]{inputenc}
\usepackage{graphicx}
\usepackage{amsmath}
\usepackage{palatino}
\usepackage{mathpazo}
\usepackage{textcomp}
\usepackage{float}
\usepackage{booktabs}
\usepackage{dcolumn}
\usepackage{ragged2e}
\usepackage{hyperref}
\hypersetup{colorlinks,citecolor=blue}
\hypersetup{colorlinks=true,linkcolor=red,filecolor=magenta,    urlcolor=blue}
\usepackage{amsmath}
\usepackage{graphicx}

\usepackage{hyperref}
\usepackage[mathlines]{lineno}
\usepackage{amssymb}
\usepackage{enumitem}
\usepackage{mathtools}
\usepackage{mathrsfs}
\usepackage{setspace}
\usepackage{color}
\usepackage{graphicx}
\usepackage[dvipsnames]{xcolor}
\usepackage{mathalfa}
\usepackage{calligra}
\usepackage[misc]{ifsym}
\DeclareMathAlphabet{\mathpzc}{OT1}{pzc}{m}{it}
\DeclareMathAlphabet{\mathcalligra}{T1}{calligra}{m}{n}
\hypersetup{colorlinks=true,citecolor=blue,linkcolor=red,urlcolor=magenta}
\usepackage{xcolor}
\usepackage{orcidlink}
\usepackage[caption=false]{subfig}
\usepackage{commath}
\def\jnl@style{}
\def\aaref@jnl#1{{\jnl@style#1}}

\def\aaref@jnl#1{{\jnl@style#1}}

\def\aj{\aaref@jnl{AJ}}                   
\def\apj{\aaref@jnl{ApJ}}                 
\def\apjl{\aaref@jnl{ApJ}}                
\def\apjs{\aaref@jnl{ApJS}}               
\def\apss{\aaref@jnl{Ap\&SS}}             
\def\aap{\aaref@jnl{A\&A}}                
\def\aapr{\aaref@jnl{A\&A~Rev.}}          
\def\aaps{\aaref@jnl{A\&AS}}              
\def\mnras{\aaref@jnl{Mon.~Not.~Roy.~Astron.~Soc.}}             
\def\prd{\aaref@jnl{Phys.~Rev.~D}}        
\def\plb{\aaref@jnl{Phys.~Lett.~B}}        
\def\prc{\aaref@jnl{Phys.~Rev.~C}}  
\def\prl{\aaref@jnl{Phys.~Rev.~Lett.}}    
\def\qjras{\aaref@jnl{QJRAS}}             
\def\skytel{\aaref@jnl{S\&T}}             
\def\ssr{\aaref@jnl{Space~Sci.~Rev.}}     
\def\zap{\aaref@jnl{ZAp}}                 
\def\nat{\aaref@jnl{Nature}}              
\def\aplett{\aaref@jnl{Astrophys.~Lett.}} 
\def\apspr{\aaref@jnl{Astrophys.~Space~Phys.~Res.}} 
\def\physrep{\aaref@jnl{Phys.~Rep.}}      
\def\physscr{\aaref@jnl{Phys.~Scr}}       
\def\commat{\aaref@jnl{Comm.~Math.~Phys.}}              
\def\science{\aaref@jnl{Science}}               
\def\cqg{\aaref@jnl{Classical Quant.~Grav.}}            
\def\jpcs{\aaref@jnl{JPCS}}                                     
\def\ijmpd{\aaref@jnl{Int.~J.~Mod.~Phys.~D}}                    
\def\grg{\aaref@jnl{Gen.~Relat.~Gravit.}}               
\def\rpp{\aaref@jnl{Rep.~Prog.~Phys.}}          
\def\npa{\aaref@jnl{Nucl.~Phys.~A}}        
\def\lrr{\aaref@jnl{Living Rev.~Rel.}}                   
\def\jcap{\aaref@jnl{J.~Cosmology Astropart.~Phys.}}    
\def\rmp{\aaref@jnl{Rev.~Mod.~Phys.}}   
\def\epjc{\aaref@jnl{Eur.~Phys.~J.~C}}

\allowdisplaybreaks[1]
\renewcommand{\arraystretch}{1.1}
\begin{document}

\preprint{APS/123-QED}

\title{Wormhole Geometry and Three-Dimensional Embedding in Extended Symmetric Teleparallel Gravity}

 
\author{V. Venkatesha\orcidlink{0000-0002-2799-2535}}%
 \email{vensmath@gmail.com (Tel: +91 9482448927)}
\affiliation{Department of P.G. Studies and Research in Mathematics,
 \\
 Kuvempu University, Shankaraghatta, Shivamogga 577451, Karnataka, INDIA
}%

\author{Chaitra Chooda Chalavadi\orcidlink{0000-0003-1976-2307}}
\email{chaitra98sdp@gmail.com} 
\affiliation{Department of P.G. Studies and Research in Mathematics,
 \\
 Kuvempu University, Shankaraghatta, Shivamogga 577451, Karnataka, INDIA
}%

\author{N. S. Kavya\orcidlink{0000-0001-8561-130X}}
\email{kavya.samak.10@gmail.com}
\affiliation{Department of P.G. Studies and Research in Mathematics,
 \\
 Kuvempu University, Shankaraghatta, Shivamogga 577451, Karnataka, INDIA
}%

\author{P. K. Sahoo\orcidlink{0000-0003-2130-8832}}
\email{pksahoo@hyderabad.bits-pilani.ac.in}
\affiliation{
 Department of Mathematics, Birla Institute of Technology and Science-Pilani,\\
 Hyderabad Campus, Hyderabad 500078, INDIA
}%


\date{\today}

\begin{abstract}
In the present manuscript, we study traversable wormhole solutions in the background of extended symmetric teleparallel gravity with matter coupling. With the anisotropic matter distribution we probe the wormhole geometry for two different gravity models. Primarily, we consider the linear model $ \mathpzc{f}( \mathcal{Q}, \mathcal{T}) = \mathcal{Q} + 2 \, \xi \, \mathcal{T}$. Firstly, we presume a logarithmic form of shape function and analyze the scenario for different redshift functions. Secondly, for a specific form of energy density, we derive a shape function and note its satisfying behavior. Next, for the non-linear model  $\mathpzc{f}( \mathcal{Q}, \mathcal{T}) = \mathcal{Q} + \alpha \,\mathcal{Q}^2 + \beta \,\mathcal{T}$ and a specific shape function we examine the wormhole solution. Further, with the aid of embedding diagrams, we interpreted the geometry of wormhole models. Finally, we conclude results. 
\begin{description}
\item[Keywords]
Wormhole, $\mathpzc{f}(\mathcal{Q},\mathcal{T})$ gravity, energy conditions.

\end{description}
\end{abstract}

\maketitle

\section{INTRODUCTION}\label{sectionI}
    In modern decades, the expansion of the universe has been illustrated by a large number of experiments on astrophysics and cosmology such as LIGO (Laser Interferometer Gravitational-Wave Observatory) \cite{gourgoulhon}, Virgo \cite{abuter}, Event Horizon Telescope (EHT) \cite{chael, akiyama}, International Gamma-Ray Astrophysics Laboratory (INTEGRAL) \cite{winkler}, Advanced Telescope for High-Energy Astrophysics (ATHENA) \cite{barcons}, Imaging X-ray Polarimetry mission (IXPE) \cite{soffitta}, Swift \cite{burrows}, CHIME \cite{amiri}. Furthermore, recent surveys such as LISA \cite{seoane}, BINGO \cite{abdalla} and SKA \cite{hall} have the potential to impose really strong boundaries on gravity theories, restricting the enormous number of theoretical proposals which have been appearing in the literature so far.
	\par The Wormholes (WHs) are fascinating compact objects in General Relativity (GR) and modified theories of gravity \cite{capozziello}. This object describes a topological bridge connecting two distinct universes or two distinct points in the same universe. The ordinary WH has the shape of a tube that is asymptotically flat from both sides of the region. Depending on the WHs structure, the throat's radius is either constant or variable. Generally, we have been classified WHs into two types namely, static and non-static. In 1916, Flamm \cite{flamm} introduced the WH concept and he built up the schwarzschild solution of isometric embedding. After that, Einstein and Rosen \cite{einstein} adopted his idea and constructed a bridge so called Einstein-Rosen bridge. Later, Wheeler and Misner \cite{minser} coined the phrase WH in 1957.
	\par A traversable WH solution to the classical Einstein field equation was put forth by Morris and Thorne \cite{morris}. These WHs violate the energy conditions (ECs), specifically the Null Energy Condition (NEC). Therefore to accomplish the traversability of wormhole in the realm of GR, a certain type of hypothetical fluild disobeying NEC is required. In the framework of GR, one of the essential conditions for the formation of a WH is the presence of an exotic matter component, due to which the energy-momentum tensor (EMT) violates the NEC \cite{visser, hochberg}. Since it was not possible to repudiate the existence of such a candidate in the framework of GR, an alternative approach was encouraged that could minimize or nullify the usage of exotic matter. Moreover, many attempts had been carried out \cite{viss, visse, kuhfitting}.
	\par Recently, modified theories of gravity have been growing interest among researchers. These theories are the geometric extension of Einstein's GR and these are used to explain the early and late time acceleration of the cosmos. The investigation of WH solutions in various modified theories of gravity is significant in Theoretical Physics. Born-field theory \cite{richarte, eiroa, shaikh}, Rastall theory \cite{moradpour}, quadratic gravity \cite{duplessis}, curvature matter coupling \cite{garcia, ditta, garc}, Einstein-Cartan gravity \cite{bronnik, bronnikov, mehdizadeh}, and Braneworld \cite{camera, lobo, kim, parsaei, kar} are some of the works that has been done on compact objects. Let's focus on WH geometry, Lobo and Oliveira \cite{oliv} studied  traversable WH in the background of $\mathpzc{f}(\mathcal{R})$ gravity. They determined the factors that were responsible for the dissatisfaction of the NEC and supported the wormhole structures. They obtained various solutions by taking various equations of state and considering some particular shape functions with constant redshift function. Saiedi and Esfahani \cite{saiedi}, using constant shape and redshift functions, obtained wormhole solutions in the framework of $\mathpzc{f}(\mathcal{R})$ gravity and investigated NEC and Weak Energy Condition (WEC). A few prominent results pertaining to the viable WHs can be seen  in \cite{kanti}. Here the discussion is made in the absence of exotic matter. In \cite{capozzi}, author derived the exact traversable WH solutions in the framework of $\mathpzc{f}(\mathcal{R})$ gravity with no exotic matter and stable conditions over the geometric fluid entering the throat. In this regard, they proposed power law models and two possible approaches for the shape function. Within the background of modified theories, Azizi \cite{azizi}, Rahaman et al \cite{rahaman}, Zubair et al \cite{zubair}, and Samanta et al \cite{samanta} have provided some significant aspects.
	\par Among the modified theories, the $\mathpzc{f}(\mathcal{Q})$ theory of gravity was introduced by Jimenez et al \cite{jime}. This theory is the curvature-free and torsion-free theory of gravity and  depends only on the non-metricity scalar. Numerous endeavors have been carried out on cosmological \cite{jimenez, frusciante, solanki} and compact objects \cite{fell, lin, hassan, wang, mustafa} (see ref. for more details). Later, the $\mathpzc{f}(\mathcal{Q},\mathcal{T})$ gravity theory \cite{yixin} is a recently proposed extension of symmetric teleparallel gravity in which the Lagrangian density is given by an arbitrary function of the non-metricity $\mathcal{Q}$ and the trace of matter-energy momentum tensor $\mathcal{T}$. This new formulated $\mathpzc{f}(\mathcal{Q},\mathcal{T})$ gravity is constructed in a similar way to the  $\mathpzc{f}(\mathcal{R}, \mathcal{T})$ \cite{harko, nojiri}, but with the geometric part of the action being replaced by the symmetric teleparallel formulation. The coupling between $\mathcal{Q}$ and $\mathcal{T}$ results in the non-conservation of the EMT, similar to the regular curvature trace of the EMT couplings. According to this theory, there is a connection between the manifestations of the quantum field due to the trace of the EMT, $\mathcal{T}$ and the gravitational effects due to the non-metricity $\mathcal{Q}$. Although, a lot of research has been done on this recently proposed gravity based on the theoretical \cite{najera, bhatta} and observational \cite{arora} aspects. Harko et al. studied the novel couplings between non-metricity and matter in \cite{koivisto}. In \cite{moreshwar},  studied the static spherically symmetric wormhole solution in $\mathpzc{f}(\mathcal{Q}, \mathcal{T})$.
	\par Motivated by the above studies, we intended to study the wormhole solution in $\mathpzc{f}(\mathcal{Q}, \mathcal{T})$ gravity for static and spherically symmetric configuration. Further, we extend our analysis by considering certain shape functions and redshift functions to find the wormhole solutions for both cases under linear and nonlinear models. The outlines of this manuscript layered as follows: Sec.\ref{sectionII} gives the criteria for traversable WH. We derive the field equations of $\mathpzc{f}(\mathcal{Q}, \mathcal{T})$ gravity in Sec.\ref{sectionIII}. In Sec.\ref{sectionIV}, we discussed the framework of the WH solutions in $\mathpzc{f}(\mathcal{Q}, \mathcal{T})$ gravity. Further, we analyze four WH solutions for a linear model in Sec.\ref{section V} and one WH solution for a non-linear model in Sec.\ref{section VI}. The geometric embedding of WH solutions has been discussed in Sec.\ref{section VII}. Ultimately, Sec.\ref{sectionVIII} gives some concluding remarks.
\section{CRITERIA FOR A TRAVERSABLE WORMHOLE}\label{sectionII}
	Consider a spherically symmetric and static Morris-Thorne WH metric \cite{morris} in the schwarzschild coordinates $(t, r, \theta, \phi)$ given by,
	
		\begin{equation}\label{whmetric}
			ds^2=-e^{2\Phi(r)}dt^2+\dfrac{dr^2}{1-\dfrac{b(r)}{r}  }  + r^2\left(d\theta^2+\text{sin}^2\theta \,d\phi^2\right),
		\end{equation}  
		where $\Phi(r)$ and $b(r)$ are both functions of radial coordinate $r$. The function $\Phi(r)$ is known as redshift function. For the WH to be traversable, one must demand that there are no horizons present, which are identified as the surfaces with $e^{2\Phi}\to 0$, so that $\Phi(r)$ must be finite everywhere. Another  function $b(r)$, represents the shape function. To explore the WH geometry, $b(r)$ must obey the following conditions:
		\begin{enumerate}[label=$\circ$,leftmargin=*]
			\setlength{\itemsep}{4pt}
			\setlength{\parskip}{4pt}
			\setlength{\parsep}{4pt}
			\item \textit{Throat condition}: The value of the function $b(r)$ at the throat is $r_0$ and hence $1-\frac{b(r)}{r}>0$ for $r>r_0.$ 
			\item \textit{Flaring-out condition:} The radial differential of the shape function, $b'(r)$ at the throat should satisfy, $b'(r_0)<1.$ 
			\item \textit{Asymptotic Flatness condition:} As $r\rightarrow \infty$, $\frac{b(r)}{r}\rightarrow 0$.
		\end{enumerate}
	In addition, for the profound interpretation of traversable WHs, we need to consider the proper radial distance function $l(r)$ given by,
		\begin{equation}
			\mathit{\mathit{l}}(r)=\pm \int_{r_0}^r \dfrac{dr}{\sqrt{\dfrac{r-b(r)}{r}}}.
		\end{equation}
		
		This function must be finite over radial coordinates. Initially, it is decreases from the upper universe to the WH's throat and then from the WH's throat to the lower universe. So, $l(r)$ obtained opposite signs in the upper and lower universes.
		
\section{THE FIELD EQUATIONS IN  $\mathpzc{f}(\mathcal{Q},\mathcal{T})$ GRAVITY }\label{sectionIII}	

		 For $\mathpzc{f}(\mathcal{Q},\mathcal{T})$gravity, the action integral of the gravitational field reads,
		\begin{equation}\label{action}
			S=\int \frac{1}{16\pi}\, \mathpzc{f}(\mathcal{Q},\mathcal{T})	\sqrt{-g}\, d^4x +\int \mathscr{L}_m \sqrt{-g}\, d^4x,
		\end{equation}
		where $\mathpzc{f}(\mathcal{Q},\mathcal{T})	$ represents the arbitrary function of non-metricity  $\mathcal{Q}$ and trace of the energy momentum tensor $\mathcal{T}$, $\mathscr{L}_m$ is the Lagrangian density corresponding to matter and \textbf{$g=det(g_{\mu\nu})$}.
                               
	   And non-metricity connection is
        defined by \cite{jime}
	    \begin{equation}
	        \mathcal{Q}_{\lambda\mu\nu} = \nabla_\lambda g_{\mu\nu},
	    \end{equation}
	    Now let us introduce the following superpotential as
	    
	    \begin{equation}\label{sp}
	         {P^\alpha} _{\mu\nu} = \frac{1}{4}\left[-\mathcal{Q}^\alpha_{\;\;\mu\nu} + 2\mathcal{Q}^{\;\;\;\alpha}_{\left(\mu\;\;\nu\right)}+ \mathcal{Q}^\alpha g_{\mu\nu} -\Tilde{\mathcal{Q}}^\alpha g_{\mu\nu} -\delta^\alpha_{(\mu}\mathcal{Q}_{\nu)}\right],
	    \end{equation}
	    and its two independent traces,
	    \begin{equation}
	        \mathcal{Q}_\alpha = \mathcal{Q}^{\;\;\;\mu}_{\alpha\;\;\mu}, \quad \Tilde{\mathcal{Q}}_\alpha = \mathcal{Q}{^\mu}_{\alpha\mu}.
	    \end{equation}
	    which enables us to define non-metricity scalar as
	    \begin{equation}
	        \mathcal{Q} = -\mathcal{Q}_{\alpha \mu \nu} P^{\alpha \mu \nu}.
		 \end{equation}
	   Varying the action \eqref{action} with respect to the metric provides the field equation as
    \begin{widetext}
	   \begin{equation}\label{fd}
		     \frac{-2}{\sqrt-g} \nabla_\alpha \left(\sqrt -g \mathpzc{f}_\mathcal{Q} P^\alpha _{\;\;\mu \nu}\right) - \frac{1}{2} g_{\mu\nu} \mathpzc{f} + \mathpzc{f}_\mathcal{T}\left( \mathcal{T}_{\mu \nu} + \Theta_{\mu \nu}\right)  - \mathpzc{f}_\mathcal{Q} \left(P_{\mu \alpha \beta} \mathcal{Q}^{\;\;\alpha \beta} _{\nu} -2\mathcal{Q}^{\alpha \beta} _{\;\;\mu} P_{\alpha \beta\nu}\right) = 8\pi \mathcal{T}_{\mu\nu},
		\end{equation}
  \end{widetext}
		Here $\mathpzc{f}_\mathcal{Q} = \frac{d\mathpzc{f}}{d\mathcal{Q}}$ and $\mathpzc{f}_\mathcal{T} = \frac{d\mathpzc{f}}{d\mathcal{T}}$.
		
		The EMT for the fluid depiction of spacetime can be expressed as
		\begin{equation}
		   \mathcal{T}_{\mu \nu} = \frac{-2}{\sqrt-g} \frac{\delta(\sqrt -g \mathscr{L}_m)}{\delta g^{\mu \nu}},
		\end{equation}
		and
		\begin{equation} \label{tm}
		    \Theta _{\mu \nu} = g^{\alpha \beta} \frac{\delta \mathcal{T}_{\alpha \beta}}{\delta g^{\mu \nu}}.
		\end{equation}

\section{WORMHOLE SOLUTIONS IN  $\mathpzc{f}(\mathcal{Q},\mathcal{T})$ GRAVITY}\label{sectionIV}
		In this study, we consider an anisotropic distribution of matter threading the wormhole described by the following stress-energy tensor $T_{\mu\nu}$
		\begin{equation}\label{energymomentumtensor}
			\mathcal{T}_{\mu\nu}=(\rho+p_t)u_\mu u_\nu-p_t\,g_{\mu\nu}+(p_r-p_t)x_{\mu}x_\nu,
		\end{equation}
		where $\rho, p_r, p_t, u_\mu$, and $x_\mu$ are the energy density, the radial pressure, the tangential pressure, the four-velocity vector, and the space-like vector respectively. For the anisotropic case,  the pressure $p_t$ will be orthogonal to $\mathfrak{x}_\mu$ and $p_r$ will be along $\mathfrak{u}_\mu$. Further we assume that $\rho, p_r and p_t$ depends on the radial coordinate.

		The trace of EMT is found to be $\mathcal{T}= \rho - p_r - 2p_t$. 
		In this manuscript, we assume that matter lagrangian $\mathscr{L}_m = -P$, with $P$ being the total pressure \cite{moraes} and hence Eq.\eqref{tm} can be reads
		\begin{equation}
		    \Theta _{\mu \nu} = -g_{\mu\nu}\, P - 2\,\mathcal{T}_{\mu\nu},
		\end{equation}
		where $P = \frac{p_r \,+\, 2 \,p_t}{3}$.
		
		Using Eq.\eqref{sp} we can determine the  non-metricity scalar $\mathcal{Q}$ for the metric \eqref{whmetric} written as
		\begin{equation}{\label{nms}}
		    \mathcal{Q} = -\frac{b}{r^2} \left[ \frac{rb'-b}{r(r-b)} + 2 \Phi'\right].
		\end{equation}
	
		 Now, for the WH metric \eqref{whmetric}, considering the anisotropic matter distribution \eqref{energymomentumtensor} and non-metricity scalar \eqref{nms} into the equations of motion \eqref{fd}, we can extract the nonzero component of the field equations \cite{moreshwar}, we get	 
		 \begin{widetext}
		 \begin{align}
		     \label{fe1}
		         \begin{split}
		             \rho = &-\dfrac{ \mathpzc{f}_{\mathcal{Q}} \mathpzc{f}_{\mathcal{T}} \left(-rb' \left(2r(r-b)\Phi' + b + 2r\right)+ 3b^2 +4r(b-r) \left(r(b-r)\Phi'' + \Phi' \left(r (b-r)\Phi' +3b-2r\right)\right)\right)}{48 \pi r^3 (r-b)\left(\mathpzc{f}_{\mathcal{T}}+ 8 \pi\right)}\\
		             &-\dfrac{24 \pi \mathpzc{f}_{\mathcal{Q}}\left(b\left(2r(b-r)\Phi'+b\right) + r(b-2r)b'\right)}{48 \pi r^3 (r-b) \left(\mathpzc{f}_{\mathcal{T}} + 8\pi\right) } 
		             \\&- \dfrac{r(b-r)\left(\mathpzc{f}_{\mathcal{T}}\left(2 \mathpzc{f}_{\mathcal{Q}\mathcal{Q}} \mathcal{Q}' \left(2r(b-r)\Phi' +b\right)+3\mathpzc{f} r^2\right)+ 48\pi b \mathpzc{f}_{\mathcal{Q}\mathcal{Q}} \mathcal{Q}'+ 24\pi \mathpzc{f} r^2\right)}{48\pi r^3 (r-b) \left(\mathpzc{f}_{\mathcal{T}} + 8\pi\right) },
		         \end{split}
		         \end{align}
           
		         \begin{align}
		         \label{fe2}
		         \begin{split}
		             p_r = & \dfrac{ \mathpzc{f}_{\mathcal{Q}} \mathpzc{f}_{\mathcal{T}} \left(rb' \left(2r(r-b)\Phi' + b + 2r\right)- 3b^2 - 4r(b-r) \left(r(b-r)\Phi'' + \Phi' \left(r (b-r)\Phi' +3b-2r\right)\right)\right)}{48 \pi r^3 (r-b)\left(\mathpzc{f}_{\mathcal{T}}+ 8 \pi\right)}\\
		             & + \dfrac{24\pi \mathpzc{f}_{\mathcal{Q}}\left(brb'- (3b-2r) \left(2r(b-r)\Phi' +b\right)\right)}{48 \pi r^3 (r-b)\left(\mathpzc{f}_{\mathcal{T}}+ 8 \pi\right)}
		             \\& - \dfrac{r(b-r)\left(\mathpzc{f}_{\mathcal{T}}\left(2 \mathpzc{f}_{\mathcal{Q}\mathcal{Q}} \mathcal{Q}' \left(2r(b-r)\Phi' +b\right)+3\mathpzc{f} r^2\right)+ 48\pi b \mathpzc{f}_{\mathcal{Q}\mathcal{Q}} \mathcal{Q}'+24\pi \mathpzc{f} r^2\right)}{48 \pi r^3 (r-b)\left(\mathpzc{f}_{\mathcal{T}}+ 8 \pi\right)}, \quad and
		         \end{split}
		         \end{align}  
           
		         \begin{align}
		         \label{fe3}
		      \begin{split}
		          p_t = & -\dfrac{ \mathpzc{f}_{\mathcal{Q}} \mathpzc{f}_{\mathcal{T}} \left(-rb' \left(2r(r-b)\Phi' + b + 2r\right)+ 3b^2 +4r(b-r) \left(r(b-r)\Phi'' + \Phi' \left(r (b-r)\Phi' +3b-2r\right)\right)\right)}{48 \pi r^3 (r-b)\left(\mathpzc{f}_{\mathcal{T}}+ 8 \pi\right)}\\
		          & - \dfrac{24\pi r\mathpzc{f}_{\mathcal{Q}} \left(r \left(2(b-r)^2\left(\Phi'' + (\Phi')^2\right) + (2r - 5b)\Phi' + b'\left(\Phi' (b-r)-1\right)\right) +b(3b\Phi' +1)\right)}{48 \pi r^3 (r-b)\left(\mathpzc{f}_{\mathcal{T}}+ 8 \pi\right)}\\
		          &- \dfrac{r(b-r)\left(\mathpzc{f}_{\mathcal{T}}\left(2b \mathpzc{f}_{\mathcal{Q}\mathcal{Q}} \mathcal{Q}'+3\mathpzc{f} r^2\right) + 4r(b-r)\mathpzc{f}_{\mathcal{Q}\mathcal{Q}} \left(\mathpzc{f}_{\mathcal{T}} +12\pi\right) \mathcal{Q}\Phi' +24 \pi \mathpzc{f} r^2\right)}{48 \pi r^3 (r-b)\left(\mathpzc{f}_{\mathcal{T}}+ 8 \pi\right)}.
		      \end{split}
		 \end{align}
		 
		 \end{widetext}
		 
	Finally, we have three independent equations \eqref{fe1}-\eqref{fe3} for our six unknown quantities i.e., $\rho, p_r, p_t, f, \Phi$ and $b$. Thus the above system of equations is under determined and it is possible to adopt different strategies to construct wormhole solutions.
		
		
		\subsection{Energy Conditions}
    ECs determine the physical phenomenon of motion of energy and matter that arise as a result of the Raychaudhuri equation. To analyze the geodesic motion, we shall consider the criterion for different ECs. For the anisotropic matter distribution \eqref{energymomentumtensor} with $\rho, p_r$ and $p_t$ being energy density, radial pressure and tangenial pressure, suppose that $n^\mu$ ia a null vector and $u^\mu$ is a timelike vector, then we have the following:
		\begin{enumerate}[label=$\circ$,leftmargin=*]
			\setlength{\itemsep}{4pt}
			\setlength{\parskip}{4pt}
			\setlength{\parsep}{4pt}
			\item \textit{Null Energy Conditions (NEC)}: Both $\rho+p_t$ and $\rho+p_r$ are non-negative.
			\item \textit{Weak Energy Conditions (WEC)}: For non-negative energy density, it implies $\rho+p_t$ and $\rho+p_r$ are both non-negative.
			\item \textit{Strong Energy Conditions (SEC)}: For non-negative $\rho+P_j$,  $\rho+\sum_j P_j$ is non-negative. 
			\item \textit{Dominant Energy Conditions (DEC)}: For non-negative energy density, it implies $\rho- |p_r|$ and $\rho-|p_t|$ are both non-negative. 
		\end{enumerate}
		
		In this present work, we analyze two forms of non-minimal coupling of $\mathpzc{f}( \mathcal{Q}, \mathcal{T})$ models, i.e., i) the linear model and ii) the non-linear model. We consider two interesting specific shape functions $b(r) = r_0\, \dfrac{\log(r + 1)}{\log(r_0 + 1)}$ , $b(r) = \frac{r_0^2}{r}$ and the specific energy density $\rho = \rho_0 \left(\dfrac{r_0}{r}\right)^n$. Now, in further study , we are going to use the ECs to check the viability of WHs.

		\section{LINEAR MODEL OF $\mathpzc{f}( \mathcal{Q}, \mathcal{T})$ GRAVITY} \label{section V}
		In this section, let us consider a linear form of $\mathpzc{f}( \mathcal{Q}, \mathcal{T})$ gravity defined by,
		\begin{align}\label{lm1}
		    \mathpzc{f}( \mathcal{Q}, \mathcal{T}) = \mathcal{Q} + 2 \, \xi \, \mathcal{T},
		\end{align}
		where $\xi$ be the arbitrary constant. When $\xi=0$, the study reduces to GR. We know that $\mathpzc{f}( \mathcal{Q}, \mathcal{T})$ gravity is constructed in a similar way to the $\mathpzc{f} (\mathcal{R}, \mathcal{T} )$ theory, but with the geometric part of the action being replaced by the symmetric teleparallel formulation. So, we consider this model based on  $\mathpzc{f}( \mathcal{R}, \mathcal{T})$ model \cite{godani}.
		
		For the model in hand, we are going to use the shape function (SF-1) as
 \subsection{  SF-1:  $b(r) = r_0\, \dfrac{\log(r + 1)}{\log(r_0 + 1)}$} \label{SF1}
		
    In the key work of Godani \cite{godani}, SF-1 was proposed to satisfy the ECs constraints for WH in $\mathpzc{f}(\mathcal{R}, \mathcal{T})$ gravity theories. Here, SF-1 is the logarithmic form of a specific shape function and $r_0 = 1$. Using this SF-1, we analyze the scenario for different redshift functions.
 \begin{figure}[b!] 
	    \centering
	    \includegraphics[width=0.9\linewidth]{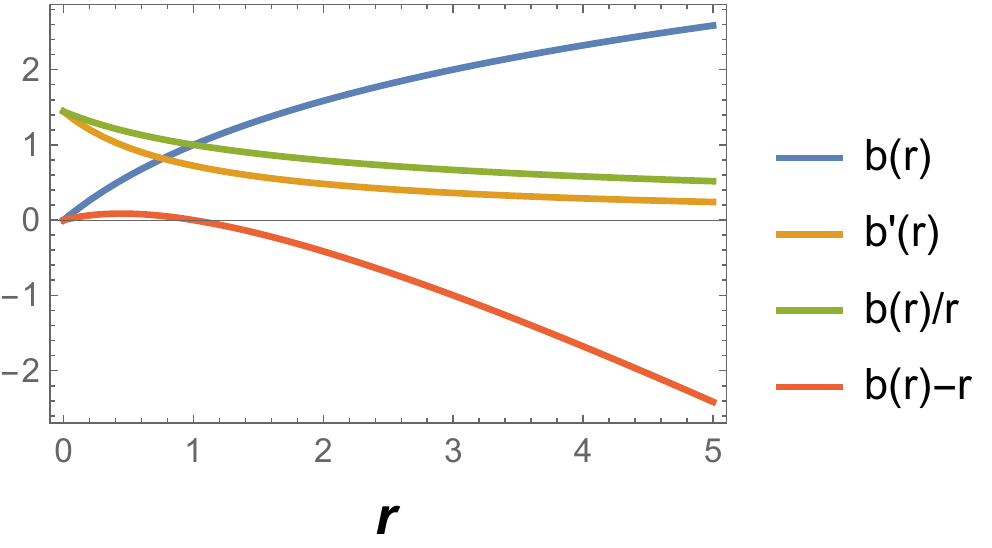}
	    \caption{Profile of shape function $b(r)$ satisfying $b'(r)<1$, $b(r)-r<0$, and $\frac{b(r)}{r}\rightarrow 0$ as $r \rightarrow \infty$.}
	    \label{fig:Asf}
	\end{figure}
	Fig.(\ref{fig:Asf}) shows the characteristics of the shape function. Clearly, one can verify that SF-1 satisfies all the necessary criteria for a traversable WH i.e., throat condition, flaring-out condition and asymptotic flatness condition.
	
      \subsubsection{\textbf{{Case}}:  $ \Phi(r) = \log {\left(1- \dfrac{1}{r} \right)}$}
      In this case, we considered the logarithmic forms of SF-1(\ref{SF1}) and redshift functions to solve the field equations \eqref{fe1}-\eqref{fe3}. After solving these equations, the energy density and radial pressure  and tangential  pressure terms are obtained which are as follows:
      
     \begin{widetext}
          \begin{align}
              \label{rho1}
               \rho = \dfrac{r \left(3 - 9\, \xi + r \left( -3+4\,\xi\right)\right) + 5 \left( 1+r\right) \xi  \log\left( 1+r\right)}{3\left(-1 +r\right) r^3 \left( 1+r\right) \left( 1+2\,\xi\right) \left(-1+ 4 \,\xi\right) \log(2)},
               \end{align}
               
               \begin{align}
                   \label{p_r1}
                   \begin{split}
         p_r = &\dfrac{r \left(-6 (1+r) \log(2) + \xi \left(-3 +24 \log(2) +8\, r \left(1 + \log(8)\right)\right)\right) }{3\left(-1 +r\right) r^3 \left(1+r\right) \left( 1+2\,\xi\right) \left(-1+ 4\,\xi\right) \log(2)}\\
         &- \dfrac{ (1+r) \left(-3+17\, \xi  +3\,r (-1+ 4 \,\xi)\right) \log( 1+r)}{3\left(-1 +r\right) r^3 \left(1+r\right) \left( 1+2\,\xi\right) \left(-1+ 4\,\xi\right) \log(2)}, \quad and
         \end{split}
               \end{align}
               
               \begin{align}
	       \label{p_t1}
	       \begin{split}
	       p_t =& \dfrac{r \left( -6 \xi \left( 1 + \log(16) \right) + \log(64) + r \left( 3 + \xi \left( 4 - 24 \log(2)\right) + \log(64)\right)\right)}{6 \left(  -1+r\right) r^3 \left( 1+r\right) \left( 1+2\,\xi\right) \left( -1+4\,\xi\right) \log(2)}\\
	       &+ \dfrac{ ( 1+r) \left( -6+14\xi + 3r( -1+4\xi)\right) \log( 1+r)}{6 \left(  -1+r\right) r^3 \left( 1+r\right) \left( 1+2\,\xi\right) \left( -1+4\,\xi\right) \log(2)}.
	       \end{split}
	   \end{align}
	
	\begin{figure*}[h]
	    \centering
	    \subfloat[Energy density $\rho$ \label{Arho}]
	    {\includegraphics[width=0.28\linewidth]{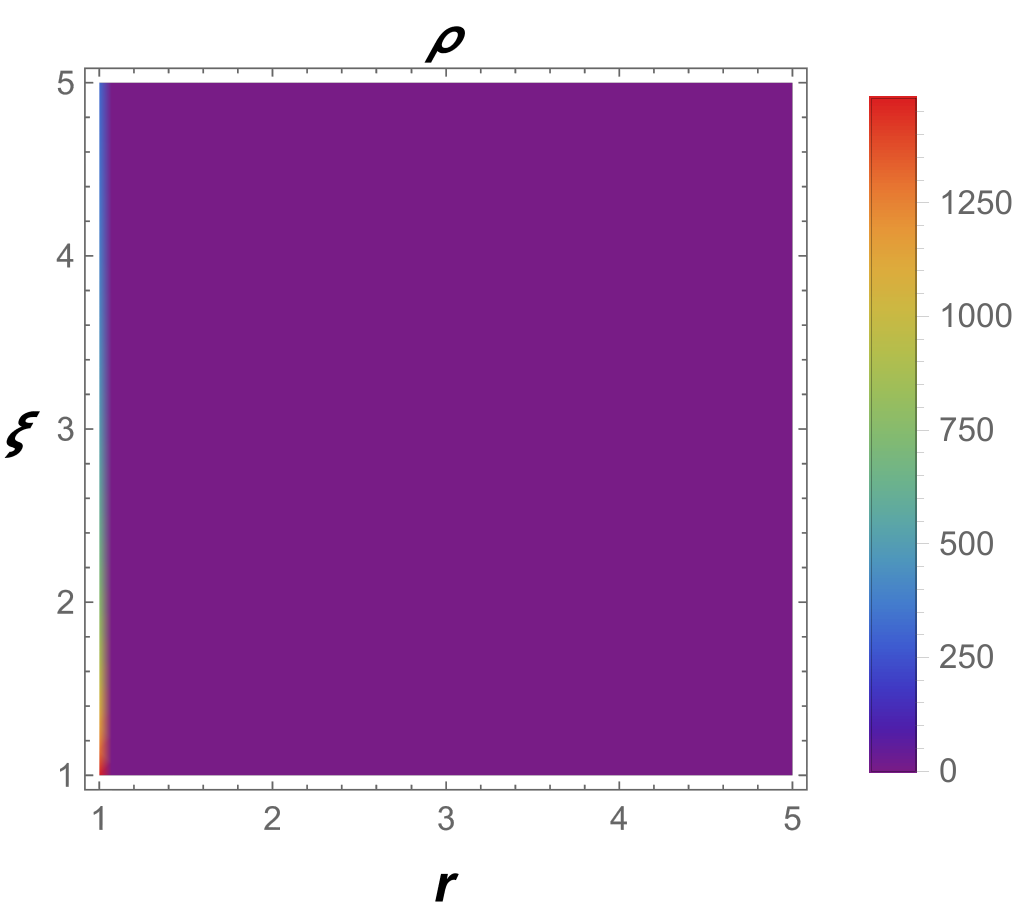}}
	    \subfloat[NEC $\rho + p_r$ \label{Aec1}]
	    {\includegraphics[width=0.28\linewidth]{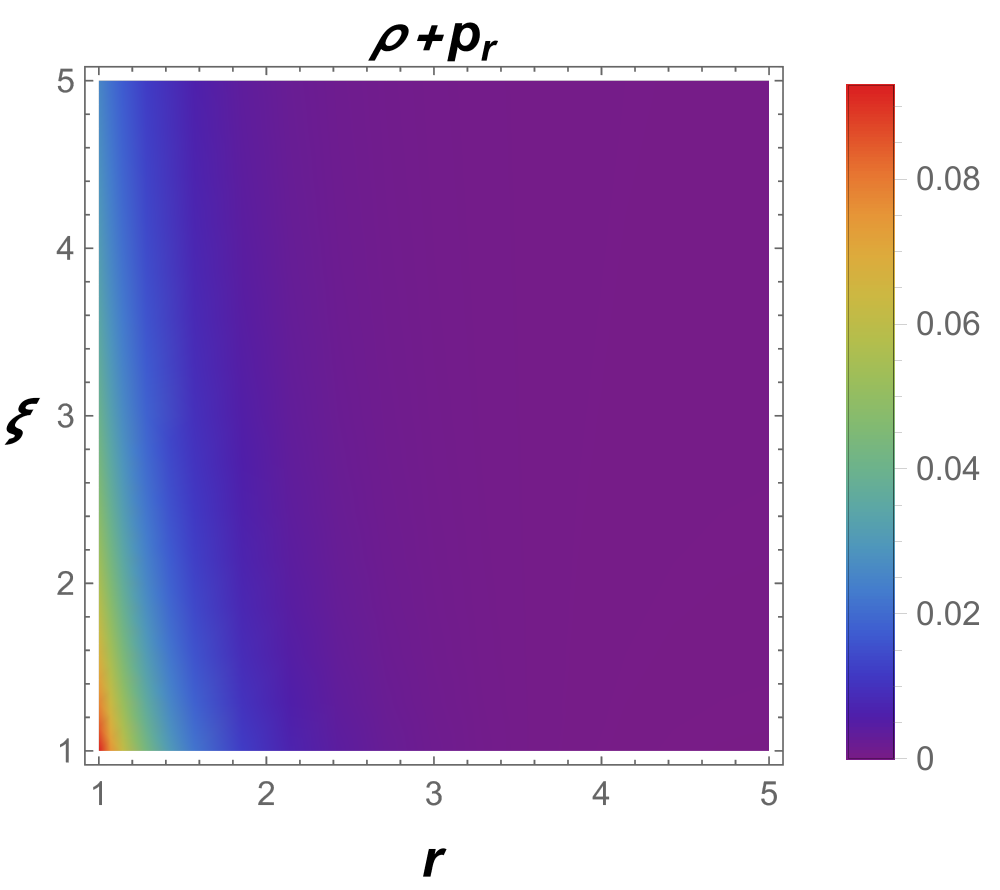}}
	     \subfloat[NEC $\rho + p_t$ \label{Aec2}]
	    {\includegraphics[width=0.27\linewidth]{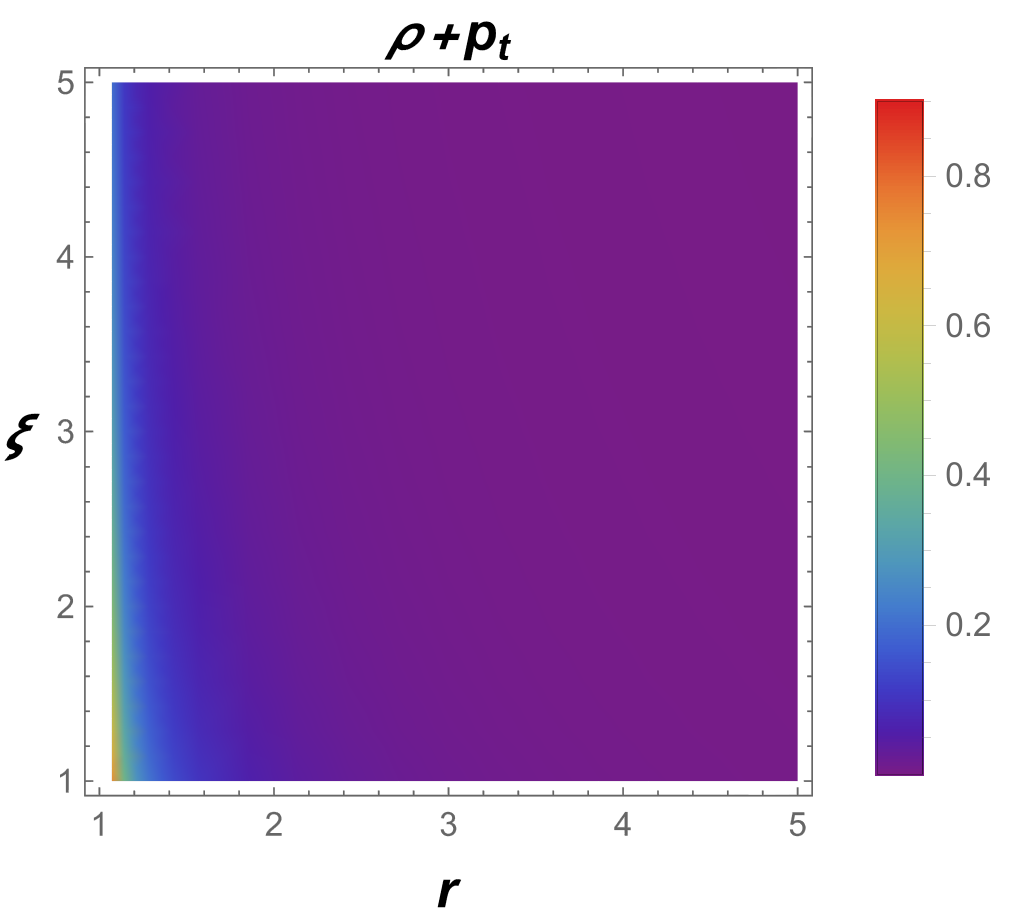}}\\
	     \subfloat[DEC $\rho -|p_r|$ \label{Aec3}]
	    {\includegraphics[width=0.28\linewidth]{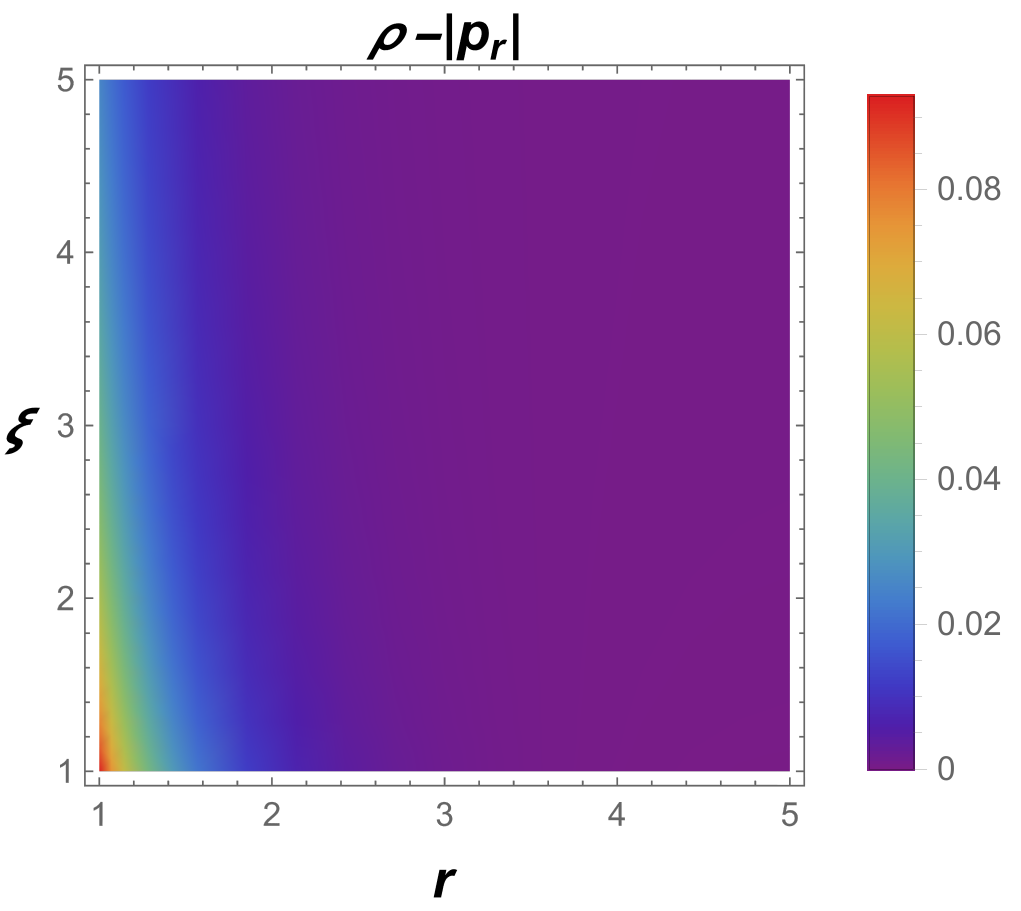}}
	     \subfloat[DEC $\rho -|p_t|$ \label{Aec4}]
	    {\includegraphics[width=0.3\linewidth]{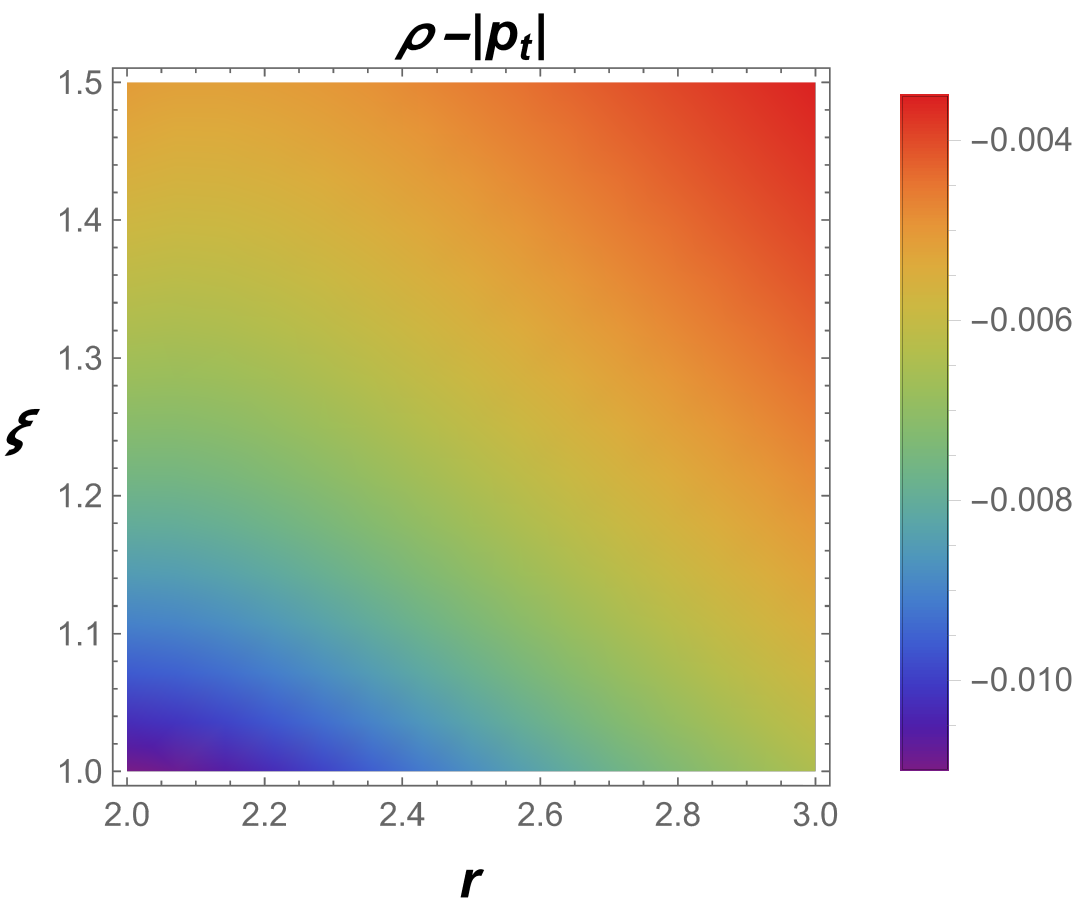}}
	     \subfloat[SEC $\rho +p_r+2p_t$ \label{Aec5}]
	    {\includegraphics[width=0.28\linewidth]{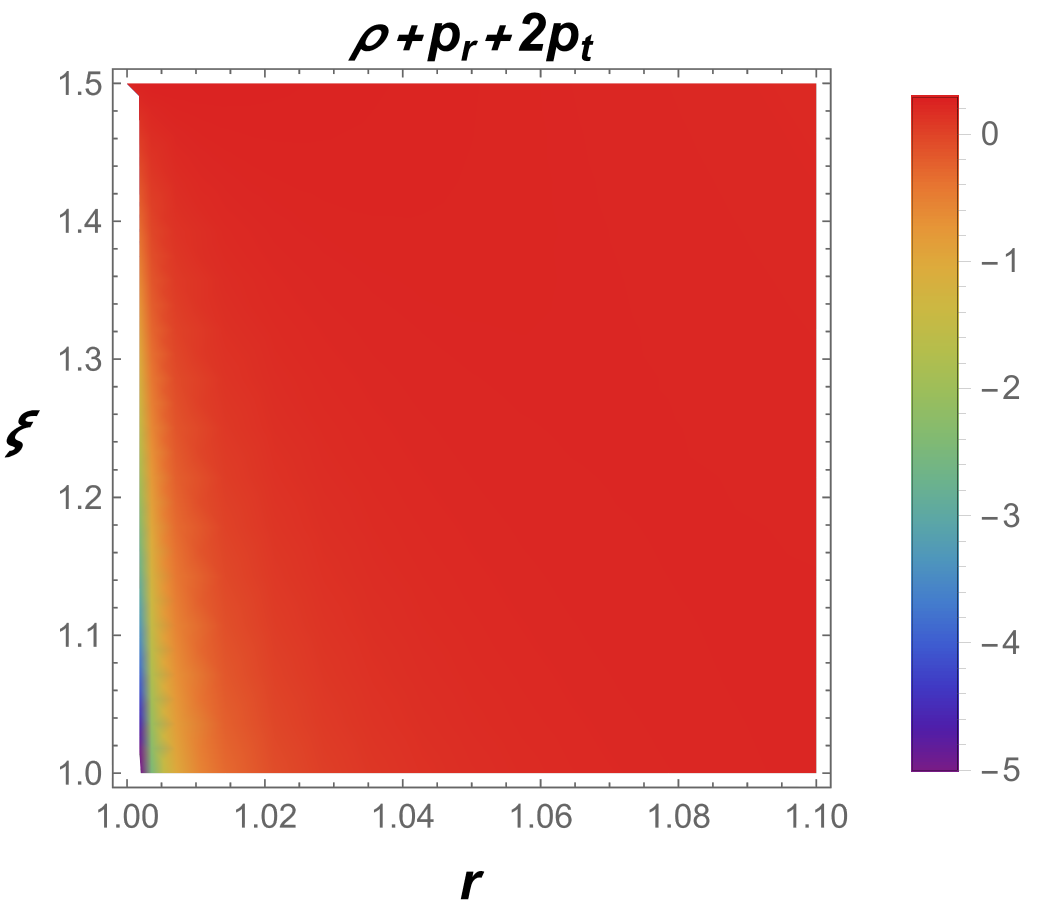}}
	    \caption{A plot depicting behaviour of energy density $\rho$, NEC, DEC, and SEC w.r.t $r$ and $\xi$ with $r_0=1$.}
	    \label{fig:Aec1}
	\end{figure*}
 \end{widetext}
		\begin{table}[!]
		\caption{The interpretation of energy conditions for case 1.}
		    \label{tab:table1}
		    \begin{center}
		    \begin{tabular}{|c|c|c|}
                \hline
				 $\xi$ & $(-\frac{1}{2},0)$  & $(\frac{1}{2},\infty)$\\ 
				\hline
				$\rho$ & obeyed & obeyed\\
				$\rho+p_r$ & obeyed & obeyed\\
				$\rho+p_t$ & obeyed & obeyed\\
				$\rho-\mid p_r\mid$ & obeyed  & obeyed\\
				$\rho-\mid p_t\mid$ & violated  & violated\\
				$\rho+p_r+2 p_t$ & obeyed & violated\\
                    \hline
			\end{tabular}
               \end{center}
		\end{table}

	 Using the above equations \eqref{rho1}-\eqref{p_t1}, we explain the graphical behavior of energy density, NEC, DEC, and SEC which are shown in Fig.(\ref{fig:Aec1}). Throughout space-time, the energy density $\rho$ is positive. From Fig.(\ref{fig:Aec1}) one can observe that NEC is satisfied. Also, there is a violation of DEC and SEC. In this case, we choose model parameter $\xi$ in  $(\frac{1}{2},\infty)$ and energy density is positive in this range. Tabel \ref{tab:table1} represents the tabulated result for the NEC, DEC, and SEC. These ECs are depicted in Fig.(\ref{fig:Aec1}).

	\subsubsection{\textbf{Case} :   $\Phi(r)= c (constant)$}
	In this section, we consider constant redshift function, so that $\Phi'(r) = 0$. The condition $\Phi'(r) = 0$ is a highly desirable feature for a traversable wormhole since it implies that the tidal forces are zero \cite{morris}. According to Kuhfitting \cite{modern, arxiv, stability}, whenever $\Phi'(r)=0$, the quantum inequality is no longer satisfied and the wormhole cannot exist on a macroscopic scale. Substituting this redshift function and SF-1(\ref{SF1}) in \eqref{fe1}-\eqref{fe3}, one can get the following energy density, radial pressure, and tangential pressure as
	\begin{widetext}
	\begin{align}
	\label{rho2}
	    \rho = \dfrac{ -3+4\,\xi}{ 3\, r^2 \left( 1+r\right) \left( 1+2\,\xi\right) \left( -1+4\,\xi\right) \log(2)},
	    \end{align}
	    
	    \begin{align}
	     \label{p_r2}
	    p_r = \dfrac{8\, r \,\xi - 3 \left( 1+r\right) \left(  -1 +4\,\xi\right) \log\left( 1+r\right)}{ 3\, r^3 \left( 1+r\right) \left( 1+2\,\xi\right) \left( -1+4\,\xi\right) \log(2)}, \quad and
	     \end{align}
	     
	     \begin{align}
	     \label{p_t2}
	    p_t = \dfrac{r (  3+4\,\xi) + 3( 1+r) ( -1+4\,\xi) \log(1+r) }{6 \,r^3 (1+r) (1+2\,\xi) ( -1+4\,\xi) \log(2)}.
	\end{align}
 
 \begin{figure*}[h!]
	    \centering
	     \subfloat[Energy density $\rho$ \label{Brho}]
	     {\includegraphics[width=0.28\linewidth]{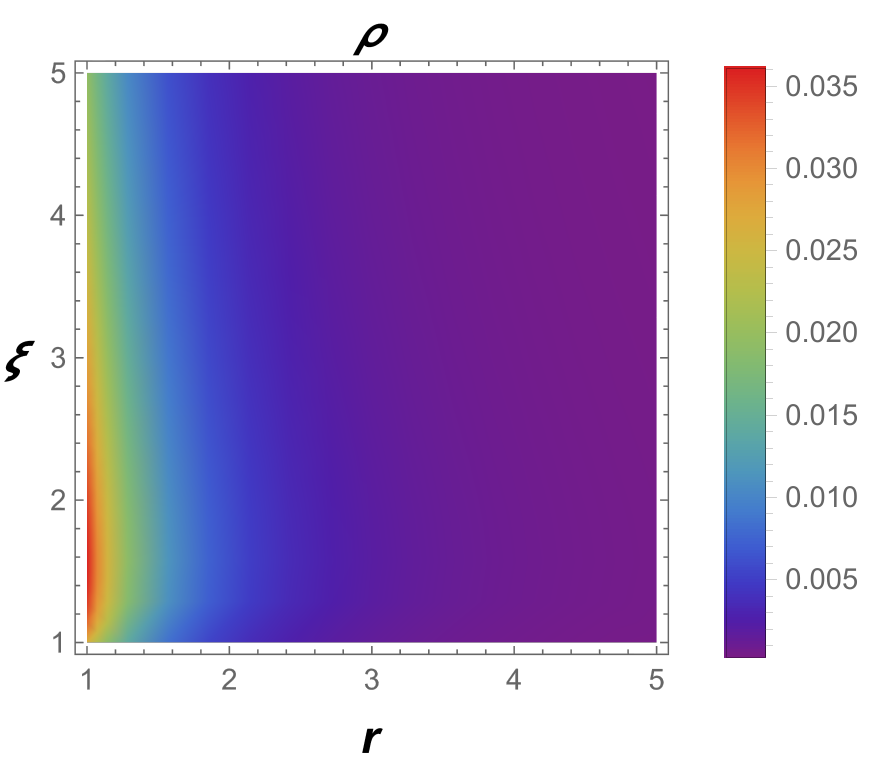}}
	      \subfloat[NEC $\rho +p_r$ \label{Bec1}]
	    {\includegraphics[width=0.28\linewidth]{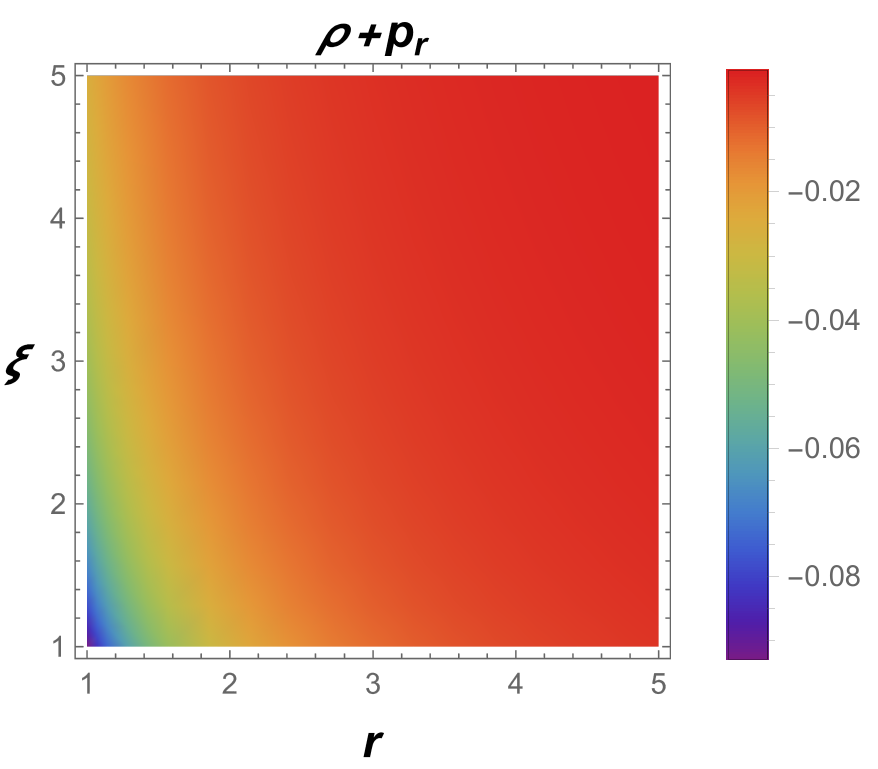}}
	    \subfloat[NEC $\rho +p_t$ \label{Bec2}]
	    {\includegraphics[width=0.28\linewidth]{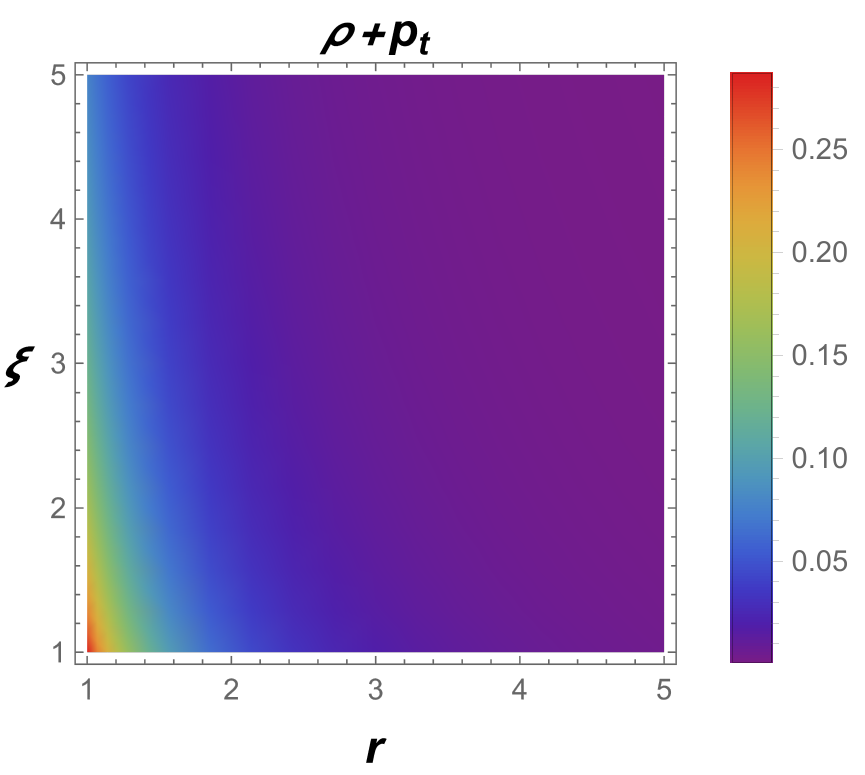}}\\
	    \subfloat[DEC $\rho-|p_r|$ \label{Bec3}]
	    {\includegraphics[width=0.28\linewidth]{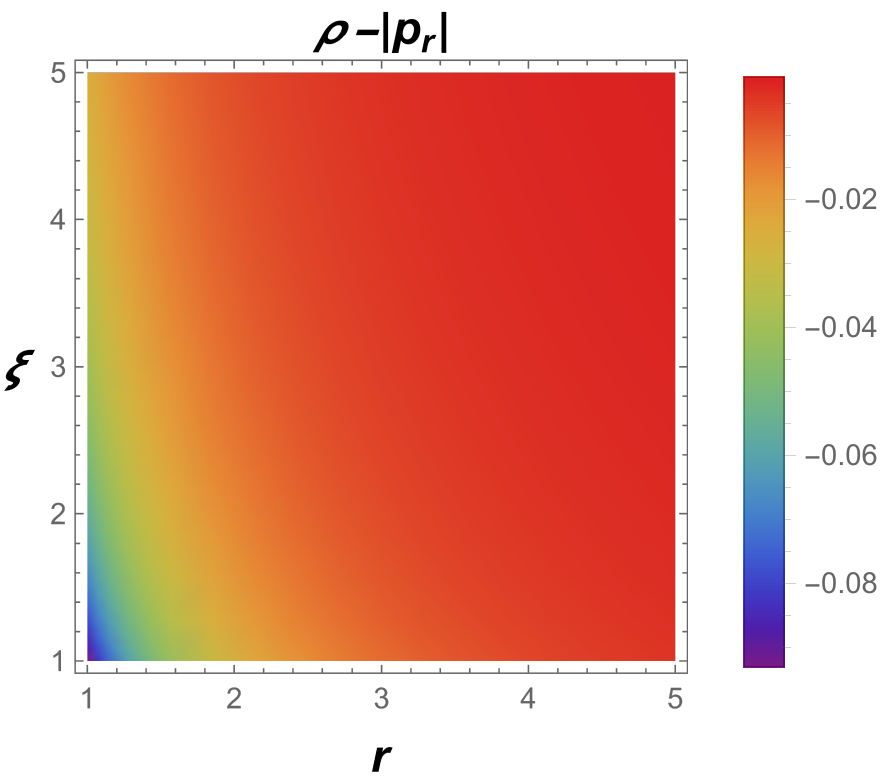}}
	    \subfloat[DEC $\rho -|p_t|$ \label{Bec4}]
	    {\includegraphics[width=0.29\linewidth]{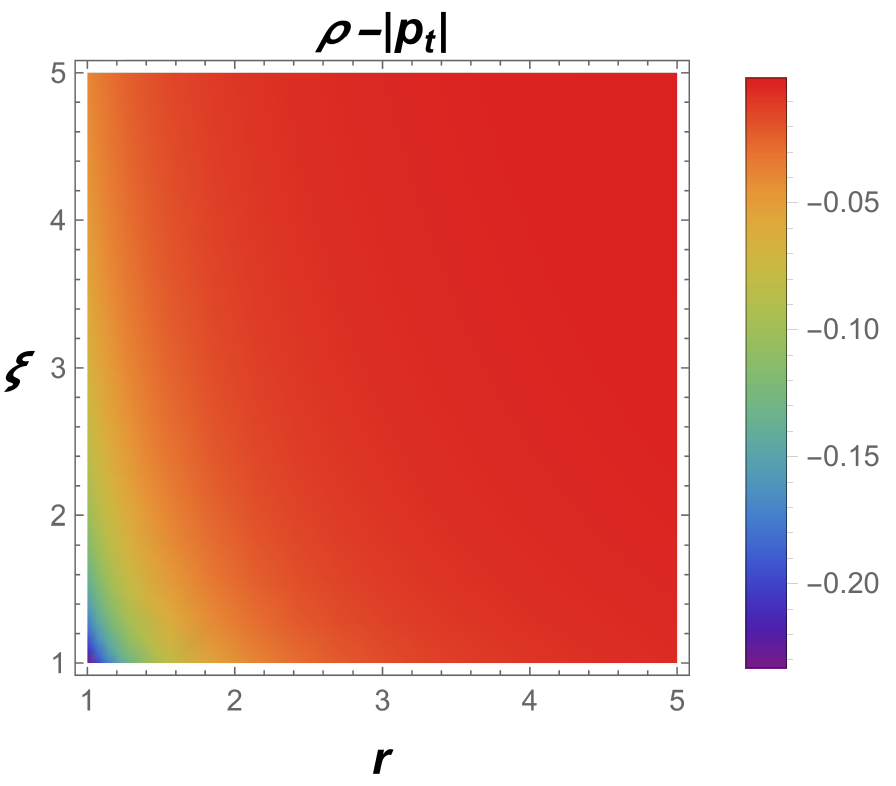}}
	    \subfloat[SEC $\rho +p_r+2p_t$ \label{Bec5}]
	    {\includegraphics[width=0.28\linewidth]{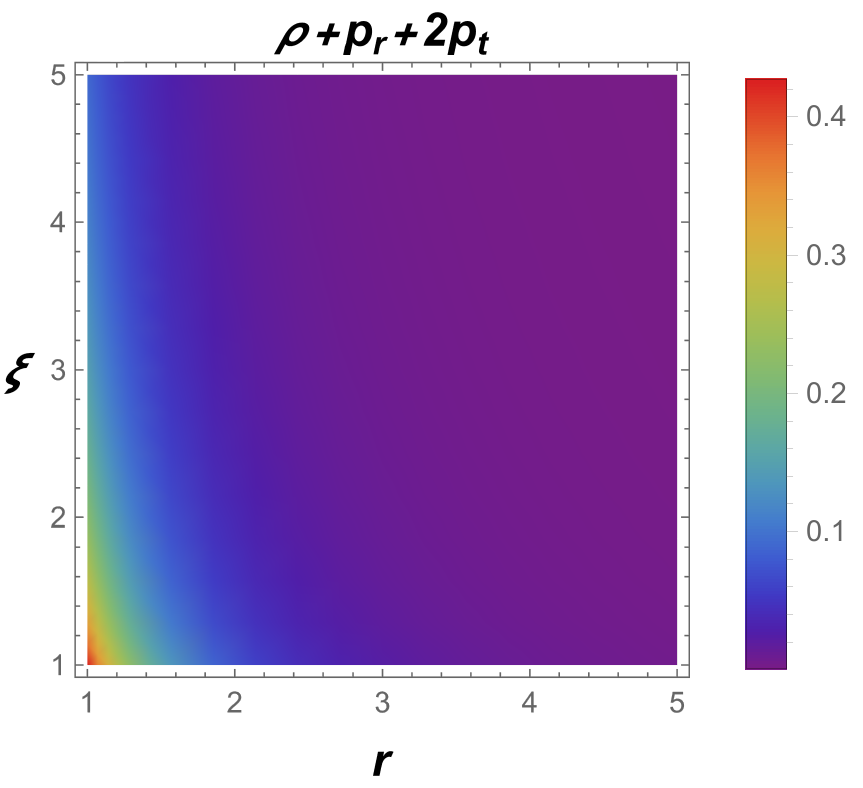}}
	    \caption{Behavior of energy density and ECs w.r.t $r$ and $\xi$ with $r_0=1$.}
	    \label{fig:ec1}
	\end{figure*}
	
	\begin{table}[b!]
		\caption{The interpretation of energy conditions for case 2.}
		    \label{tab:table2}
                 \begin{center}
		    \begin{tabular}{|c |c| c|}
      \hline
				 $\xi$ & $(-\frac{1}{2}, 0)$  & $(\frac{3}{4},\infty)$\\ 
				\hline
				$\rho$ & obeyed  & obeyed\\
				$\rho+p_r$ & violated & violated\\
				$\rho+p_t$ & obeyed & obeyed\\
				$\rho-\mid p_r\mid$ & violated & violated\\
				$\rho-\mid p_t\mid$ & violated & violated\\
				$\rho+p_r+2 p_t$ & violated & obeyed\\
    \hline
			\end{tabular}
		   \end{center}
		\end{table}	
  \end{widetext}
To verify the profiles of energy density $\rho$ and ECs have been shown in Fig.(\ref{fig:ec1}).  We know that the energy density has to be positive everywhere. From Fig.(\ref{fig:ec1}) it can be seen that the violation of NEC and DEC. Table \ref{tab:table2} explain the behavior of ECs. Similarly to the previous section we can check the valid range for the model parameter using the energy density. 

	\subsubsection{\textbf{Case} : $\Phi(r)=\frac{1}{r}$}
	Here, SF-1(\ref{SF1}) and given redshift function are substitute into the field equations \eqref{fe1}-\eqref{fe3}, we get the following equations 
 \begin{widetext}
	\begin{align}
	    \label{rho3}
	    \rho &= \dfrac{r \left(r^2 \left(-3+4\, \xi\right) + 10 \,\xi \log(2) + 5\,r \,\xi \left( 1 + \log(4)\right)\right)-5 \left( 2+ 3\,r + r^2\right) \xi \log\left(1+r\right)}{3\,r^5\left(1+r\right) \left(1+2\,\xi\right) \left(-1+4\,\xi\right) \log(2)},
	    \end{align}
	    
	    \begin{align}
	    \begin{split}
	      p_r =& \dfrac{r \left( -10\,\xi \log(2) + r \left( -5\,\xi-34\,\xi\log(2)+ \log(64)\right) +r^2\left(8\,\xi-24\,\xi\log(2)+\log(64)\right)\right)}{3\,r^5\left(1+r\right) \left(1+2\,\xi\right) \left(-1+4\,\xi\right) \log(2)} \\
	      & - \dfrac{(1+r) \left(r\left(6-29\,\xi\right)-10\,\xi+3\,r^2\left(-1+4\,\xi\right)\right)\log\left(1+r\right)}{3\,r^5\left(1+r\right) \left(1-2\,\xi\right) \left(-1+4\,\xi\right) \log(2)}, \quad and
	      \end{split}
	      \end{align}
	      
	      \begin{align}
	     \begin{split}
	      p_t = &\dfrac{r\left( 2\left(-3+2\,\xi\right)\log(2)+ r \left(\xi\left(2+28\log(2)\right)-3\left(1+\log(16)\right)\right)+r^2\left(3-6\log(2)+4\,\xi\left(1+\log(64)\right)\right)\right)}{6\,r^5 \left(1+r\right) \left(1+2\,\xi\right) \left(-1+4\,\xi\right)\log(2)}\\
	       & + \dfrac{\left(1+r\right)\left(6+r\left(9-26\,\xi\right)-4\,\xi+3\,r^2\left(-1+4\xi \right) \right) \log\left(1+r\right)}{6\,r^5 \left(1+r\right) \left(1+2\,\xi\right) \left(-1+4\,\xi\right)\log(2)}.
\end{split}
	  	\end{align}

    \begin{figure*}[h!]
	    \centering
	    \subfloat[Energy density $\rho$ \label{Case3}]
	    {\includegraphics[width=0.26\linewidth]{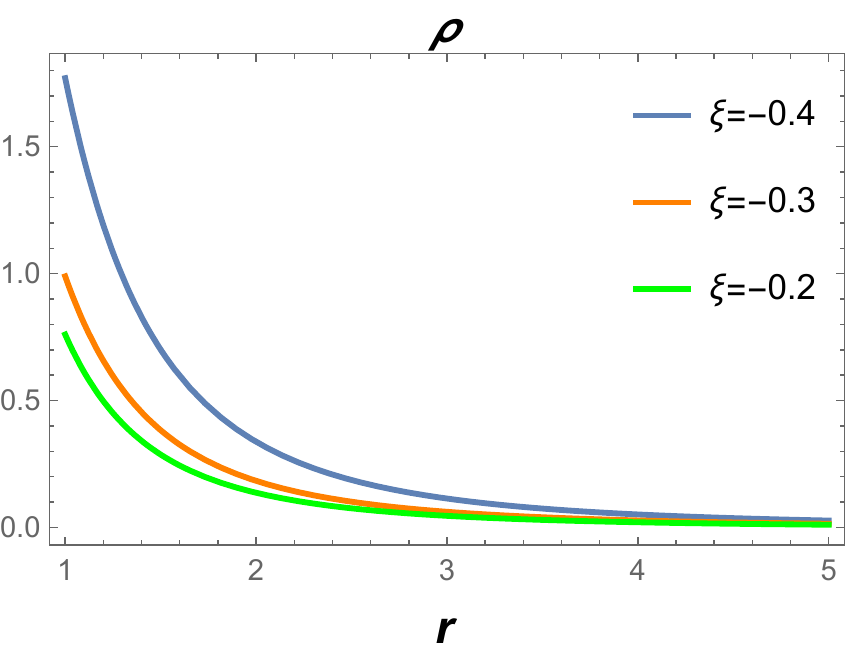}}
	    \subfloat[NEC $\rho +p_r$ \label{Cec1}]
	    {\includegraphics[width=0.265\linewidth]{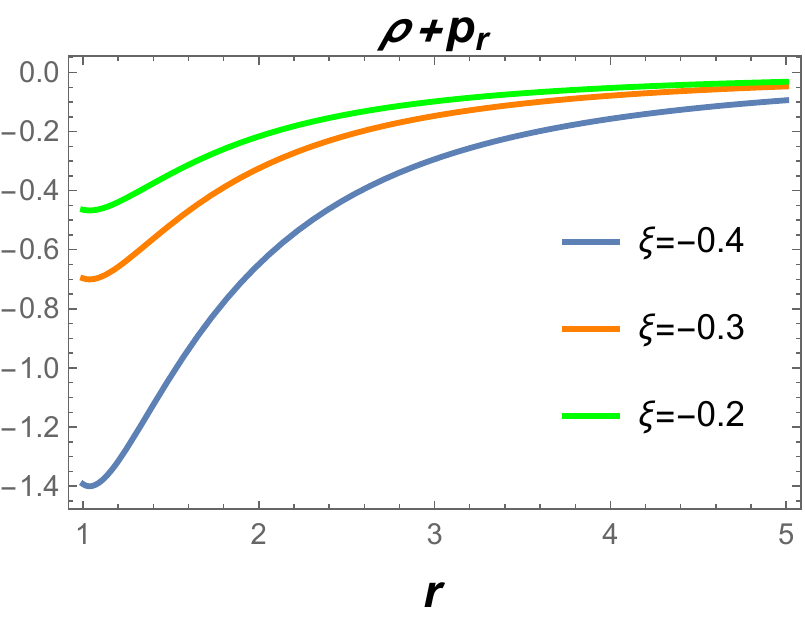}}
	    \subfloat[NEC $\rho +p_t$ \label{Cec2}]
	    {\includegraphics[width=0.26\linewidth]{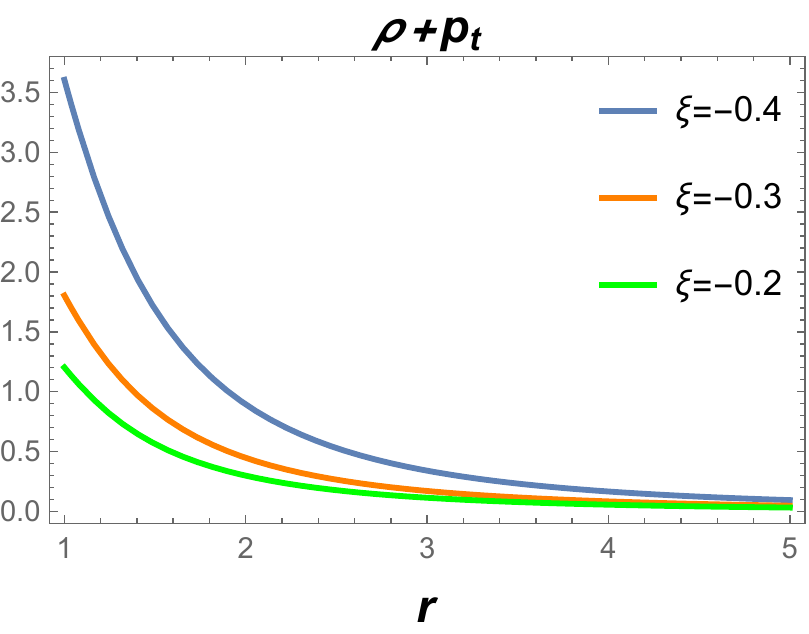}}\\
	    \subfloat[DEC $\rho -|p_r|$ \label{Cec3}]
	    {\includegraphics[width=0.26\linewidth]{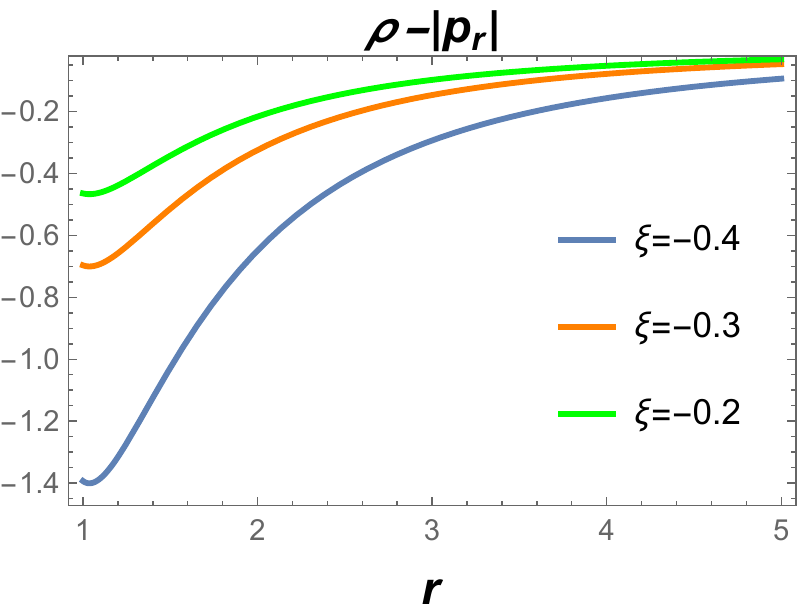}}
	    \subfloat[DEC $\rho -|p_t|$ \label{Cec4}]
	    {\includegraphics[width=0.265\linewidth]{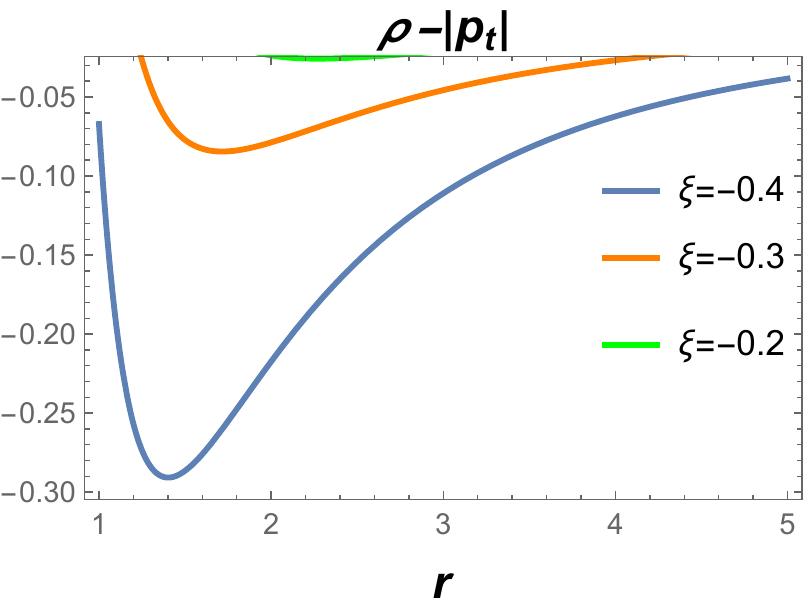}}
	    \subfloat[SEC $\rho +p_r+2p_t$ \label{Cec5}]
	    {\includegraphics[width=0.258\linewidth]{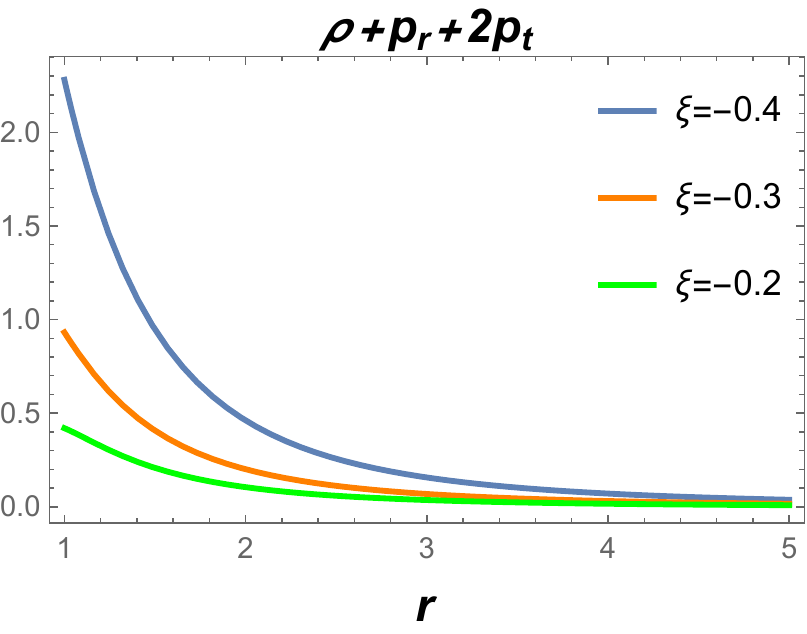}}
	    \caption{The plots of energy density and ECs with $\xi=-0.4$ (Grey), $\xi=-0.3$ (Orange), $\xi=-0.2$ (Green) and $r_0=1$.}
	    \label{fig:Cec1}
	\end{figure*}
	\end{widetext}
 In Fig.(\ref{fig:Cec1}) we have shown the graph of energy density $\rho$, which is positive throughout space-time and explains the graph of ECs for some points such as $\xi=-0.4$, $\xi=-0.3$, and $\xi=-0.2$. Moreover, NEC and DEC are violated and SEC is satisfied. These ECs has shown in Fig.(\ref{fig:Cec1}).
 
    \subsection{\textbf{Specific Energy Density (SED)}: $\rho = \rho_0 \left(\dfrac{r_0}{r}\right)^n$ and  $\Phi(r)= c$ (constant)}
    \par In the present case, we shall consider the scenario in which the energy density takes the form \cite{swkim}
    \begin{equation} \label{rho}
         \rho = \rho_0 \left(\dfrac{r_0}{r}\right)^n,
    \end{equation}
    where $\rho_0$, $n$ are constants and $r_0$ is the throat radius. Also, $ r$ lies in the region $r_0 < r <\infty$. Since the minimum value of $r=r_0>0$ in all our models. We do not have a singularity. The power law of energy density contradicts the condition for having zero perturbations propagating the fluid without letting sound speed to vanish. In \cite{capozzi}, it appears also singular at $r = 0$. In particular, the WH solutions is unstable if the sound speed does not disappear. Solving equations \eqref{fe2} and \eqref{fe3} we get the pressure elements as,
    
	\begin{eqnarray}
	p_r=\frac{5 \xi  r b'-3b (3 \xi -1)+3 \xi  \rho_0 r^3 (2\xi+1) \left(\frac{r_0}{r}\right)^n }{3 (2 \xi +1) (3 \xi -1) r^3},\\
	p_t=\frac{\xi  r b'+3 r b'+3b(3 \xi -1)+6 \xi  \rho_0 r^3(2 \xi +1) \left(\frac{r_0}{r}\right)^n}{6 (2 \xi +1) (3 \xi -1) r^3}.
	\end{eqnarray}
	
	Using the aforesaid expressions for pressure components and specific form of shape function \eqref{rho} in \eqref{fe1} we can derive the shape function (SF-2) as,
	\begin{equation}
	    b(r)=k+\frac{3 \left(8 \xi ^2+2 \xi -1\right) \rho_0 r^3 \left(\frac{r_0}{r}\right)^n}{(3-n) (4 \xi -3)},
	\end{equation}
 \begin{widetext}
	\begin{figure*}[!]
	    \subfloat[$b(r)$ \label{fig:Rsf1}]{\includegraphics[width=0.4\linewidth]{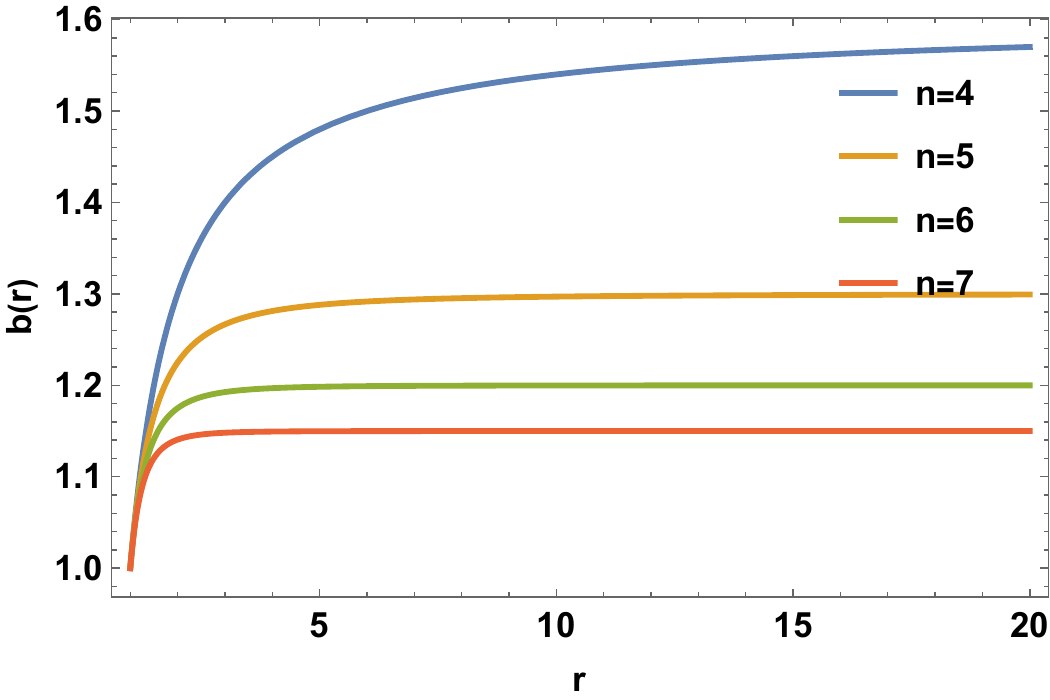}}
	    \subfloat[$b'(r)$\label{fig:Rsf2}]{\includegraphics[width=0.415\linewidth]{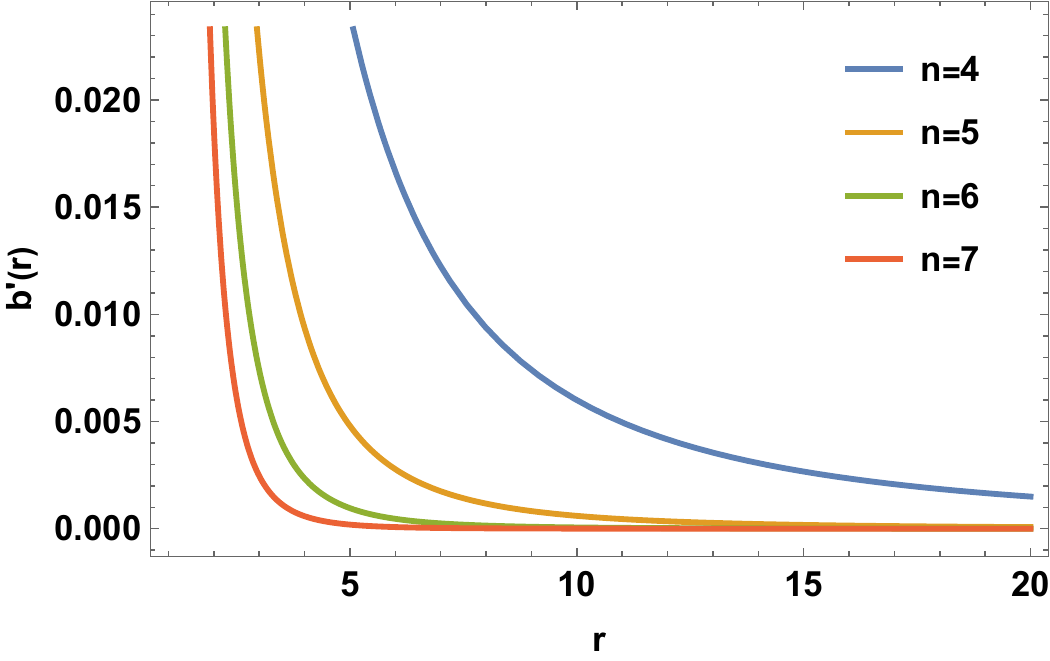}}\\
	    \subfloat[$b(r)/r$\label{fig:Rsf3}]{\includegraphics[width=0.4\linewidth]{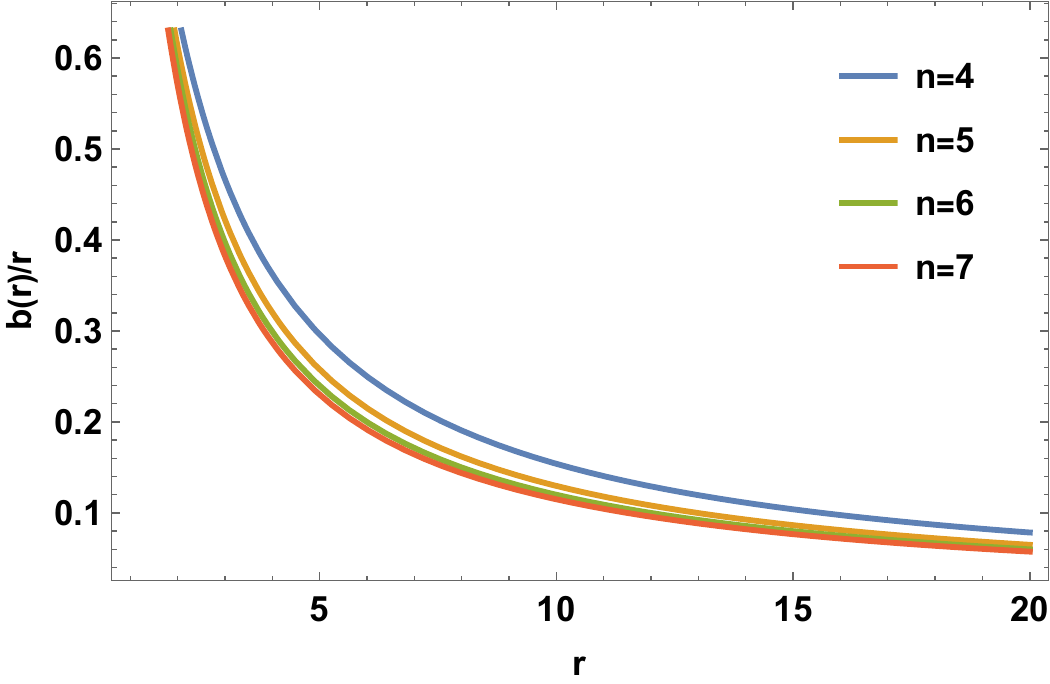}}
	    \subfloat[$b(r)-r$\label{fig:Rsf4}]{\includegraphics[width=0.41\linewidth]{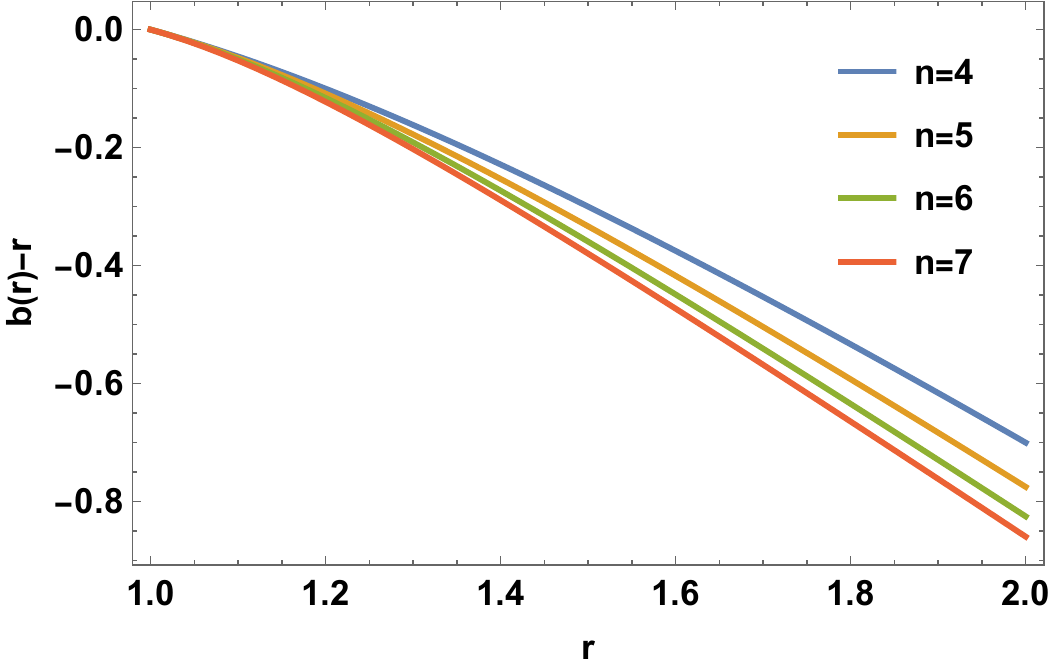}}
	    \caption{Profile of shape function $b(r)$ satisfying $b'(r)<1$, $b(r)-r<0$, and $\frac{b(r)}{r}\rightarrow 0$ as $r \rightarrow \infty$ for $r_0=1$ and $ \xi=-0.25$.}
	    \label{fig:Rsf}
	   \end{figure*}
     \end{widetext}

     \begin{widetext}
    \begin{figure*}[!]
	    \centering
	    \subfloat[Energy density $\rho$\label{fig:Rrho}]{\includegraphics[width=0.28\linewidth]{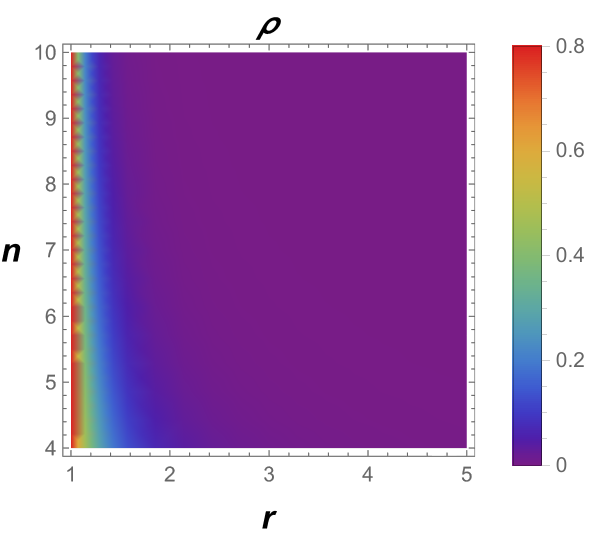}}
	    \subfloat[NEC $\rho+p_r$\label{fig:Re1}]{\includegraphics[width=0.29\linewidth]{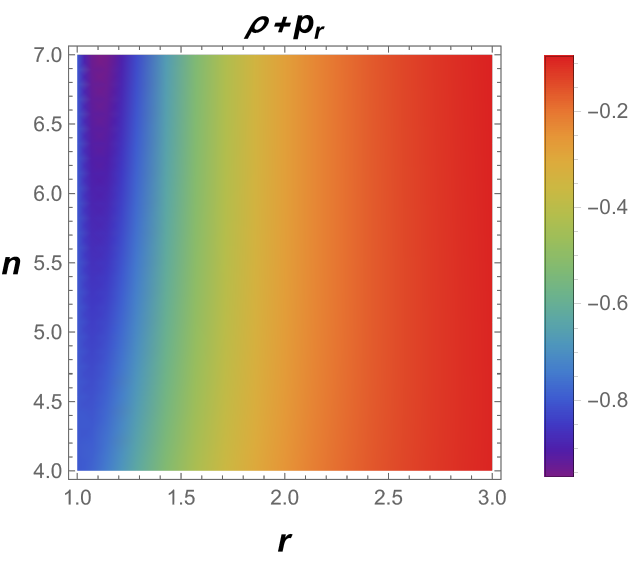}}
	    \subfloat[NEC $\rho+p_\tau$\label{fig:Re2}]{\includegraphics[width=0.285\linewidth]{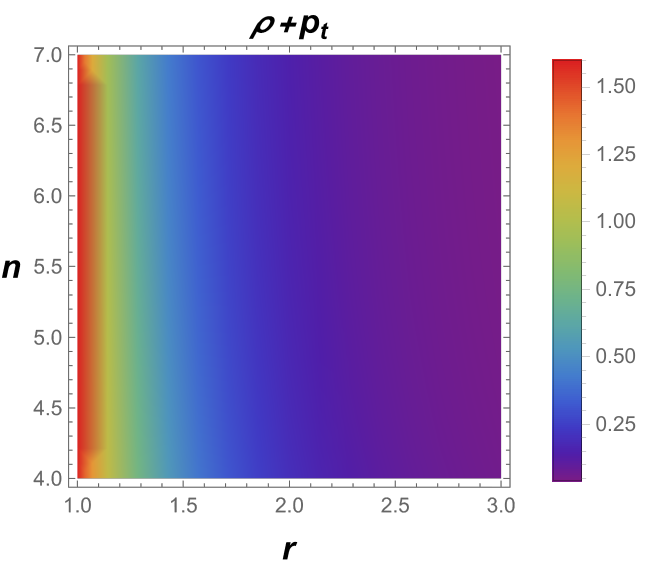}}\\
	    \subfloat[DEC $\rho-|p_r|$\label{fig:Re3}]{\includegraphics[width=0.285\linewidth]{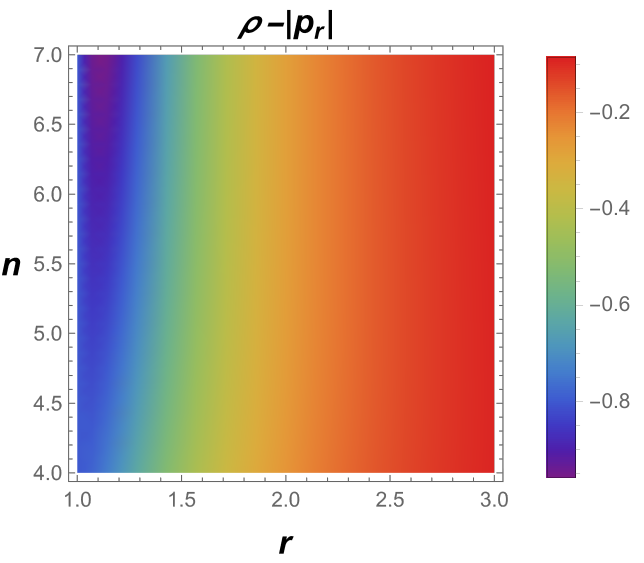}}
	    \subfloat[DEC $\rho-|p_\tau|$\label{fig:Re4}]{\includegraphics[width=0.29\linewidth]{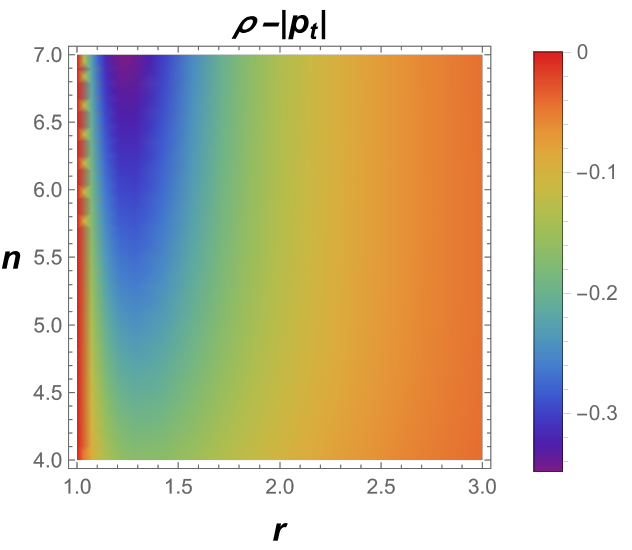}}
	    \subfloat[SEC $\rho+p_r+2p_\tau$\label{fig:Re5}]{\includegraphics[width=0.28\linewidth]{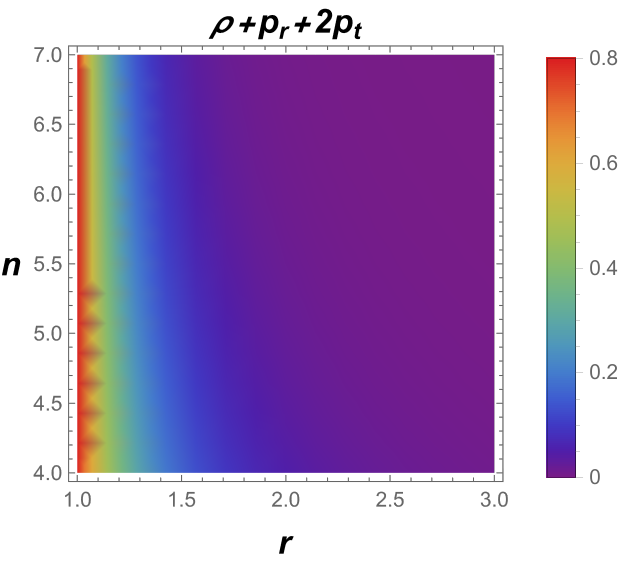}}
	    \caption{Profile of energy density and ECs w.r.t $r$ and $n$ with $r_0=1$ and $\xi=-0.25$. }
	    \label{fig:Rec}
	\end{figure*}
\end{widetext}
	where k is integrating constant and it should take the form 
		
\begin{widetext}
	\begin{equation}
    \begin{split}
        k=\frac{1}{(n-3) (4 \xi -3)}\left[4 n \xi  r_0-3 n r_0+24 \xi ^2 \rho_0 r_0^3+6 \xi  \rho_0 r_0^3-3 \rho_0 r_0^3-12 \xi  r_0+9 r_0\right].
    \end{split}
	\end{equation}
\end{widetext}

	To obey the throat condition. Further, from Fig.(\ref{fig:Rsf}) we can observe that the obtained shape function satisfies flaring-out condition and asymptotic flatness conditions. In addition, we verified the energy conditions (Fig. \ref{fig:Rec}) for the present case and observed the violation of NEC and DEC. The violation of NEC suggests the requirement of exotic matter at the wormhole throat.

	 \section{Non-linear model of $\mathpzc{f}( \mathcal{Q}, \mathcal{T})$ gravity} \label{section VI}
	Analogous to GR, we consider a non-linear form of $\mathpzc{f}( \mathcal{Q}, \mathcal{T})$ gravity given by,
      	
	\begin{align}
	    \mathpzc{f}( \mathcal{Q}, \mathcal{T}) = \mathcal{Q} + \alpha \,\mathcal{Q}^2 + \beta \,\mathcal{T},
	\end{align}
 
	where $\alpha$ and $\beta$ are model parameters. When $\alpha=\beta=0$, the study reduces to GR. This form of $\mathpzc{f}( \mathcal{Q}, \mathcal{T})$ gravity has been studied in \cite{raja}.
	
	 Here, we consider the specific shape function (SF-3) as
	
	\subsection{SF-3: $b(r) = \frac{r_0^2}{r}$}{\label{SF-3}}
     This function was proposed in the original work of Lobo et al. \cite{oliv}, where it was used to confirm the viability of WH solutions for various $\mathpzc{f}(\mathcal{R})$ gravity theories. 
	\begin{figure}[h] 
	    \centering
	    \includegraphics[width=0.9\linewidth]{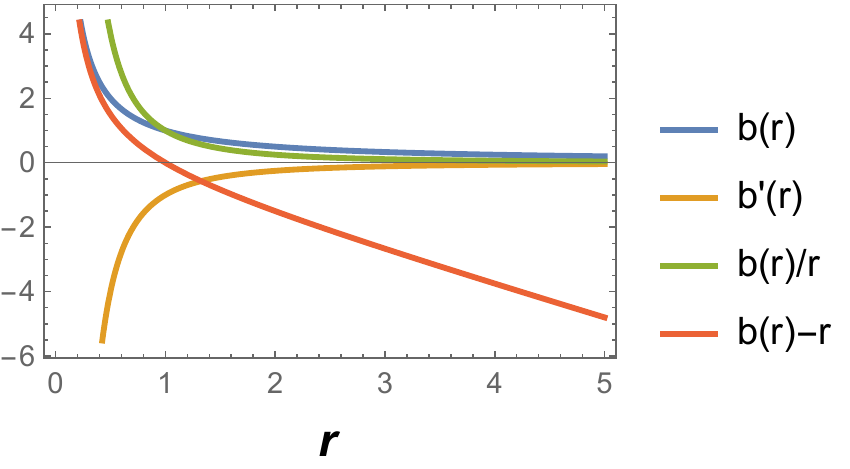}
	    \caption{A plot showing characteristics of shape function with $r_0=1$.}
	    \label{fig:Dsf}
	\end{figure}
	Now, the SF-3 (\ref{SF-3}) satisfies all necessary criteria such as throat condition, flaring-out condition and asymptotically flatness condition. These conditions are represented in Fig.(\ref{fig:Dsf}).
	\subsubsection{\textbf{Case}:  $\Phi(r) = c$ (constant) }
	
	For instance, we rely on the absence of tidal force i.e., $\Phi(r)$ considered to be constant \cite{godani,anchor,falco,modern}. So that $\Phi'(r)= 0$ which simplifies the calculations considerably, and provide interesting exact wormhole solutions. From SF-3 (\ref{SF-3}) and redshift function in the field equations \eqref{fe1}-\eqref{fe3} becomes,
	
	\begin{widetext}
	      	\begin{align}
	    \rho =\frac{ (3-2 \,\beta) (r^4 + r^8 + 28\,r^2 \,\alpha ) + r^6 ( -6 + 4 \,\beta) + 2 \alpha (  -27+23\, \beta )}{3\, r^8\, (-1+r^2)^2 \,(  -1+\beta +2 \,\beta^2)},
	\end{align}
	
	\begin{align}
	    p_r = \frac{(3 - 10 \,\beta ) (r^4 + r^8) + 4\, r^2\, \alpha ( -15 + 2 \,\beta ) + r^6 ( -6+20 \,\beta) + 2 \alpha ( 15+\beta)}{3\, r^8\, (-1+r^2)^2 \,(  -1+\beta +2 \,\beta^2)}, \quad and
	\end{align}
	
	\begin{align}
	    p_t = \frac{( -3+2\, \beta  ) ( r^4 + r^8) - 4 \,r^2\, \alpha (  3+22 \,\beta) + r^6 ( 6 - 4\,\beta) + 2\,\alpha (3 +25 \,\beta)}{3\, r^8\, (-1+r^2)^2 \,(  -1+\beta +2 \,\beta^2)}.
	\end{align}

\begin{figure*}[t!]
      \subfloat[Energy density $\rho$\label{fig:Drho}]
      {\includegraphics[width=0.3\linewidth]{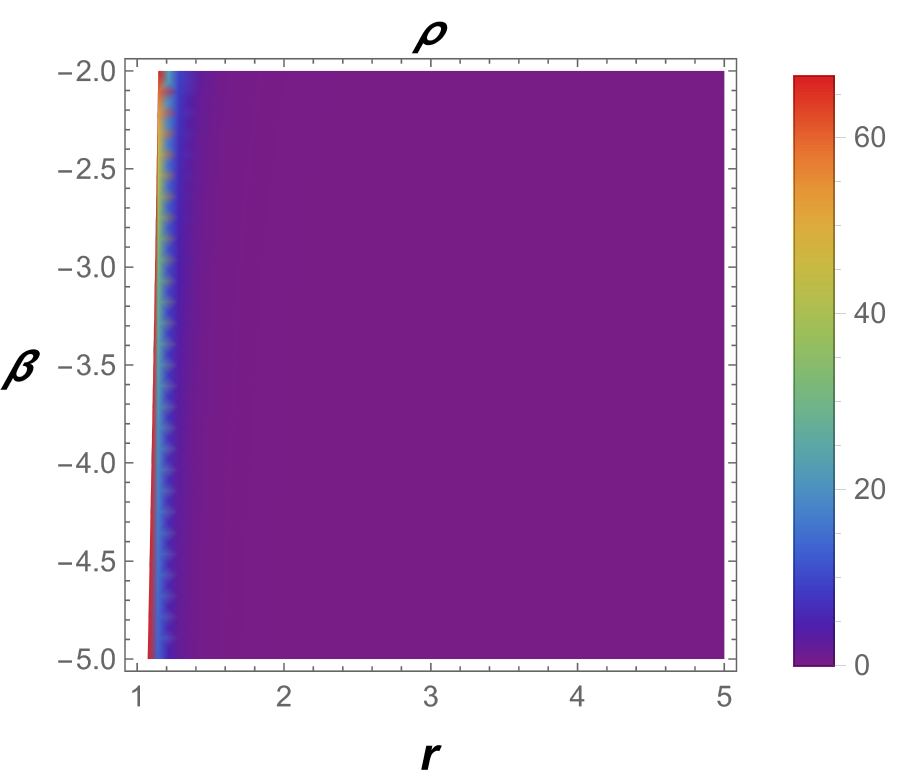}}
      \subfloat[NEC $\rho + p_r$ \label{fig:Dec1}] {\includegraphics[width=0.3\linewidth]{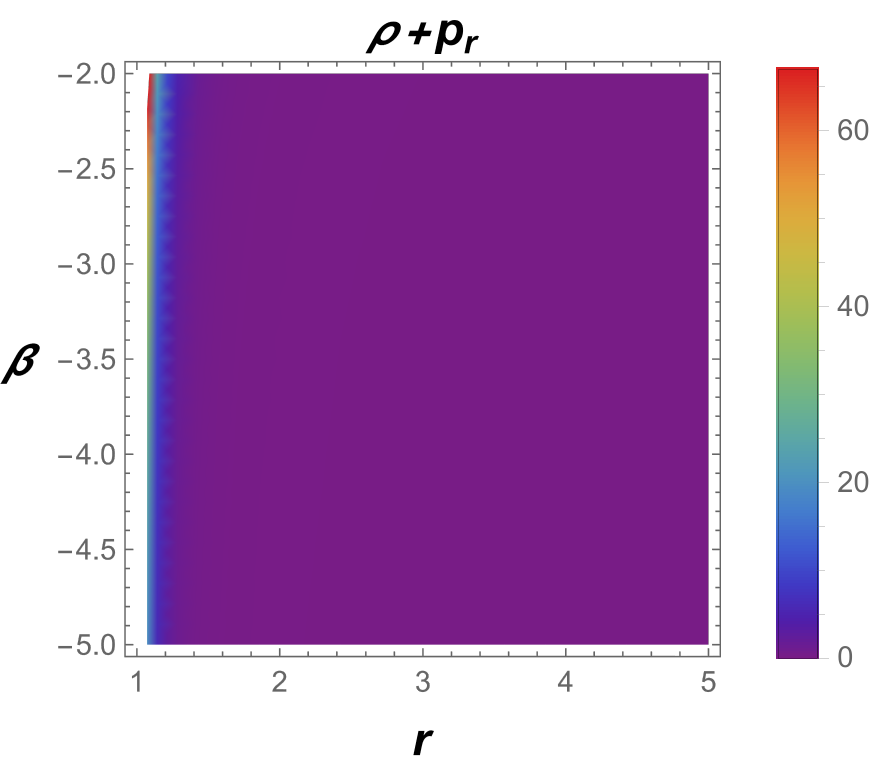}}
      \subfloat[NEC $\rho + p_t$ \label{fig:Dec2}] {\includegraphics[width=0.3\linewidth]{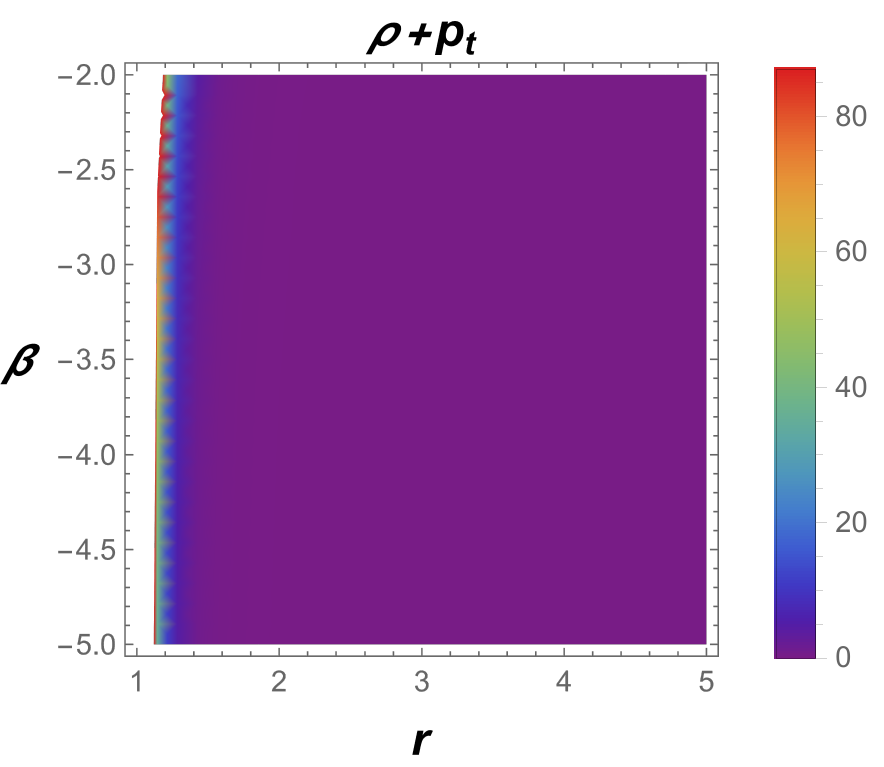}}\\
      \subfloat[DEC $\rho -| p_r|$ \label{fig:Dec3}] {\includegraphics[width=0.3\linewidth]{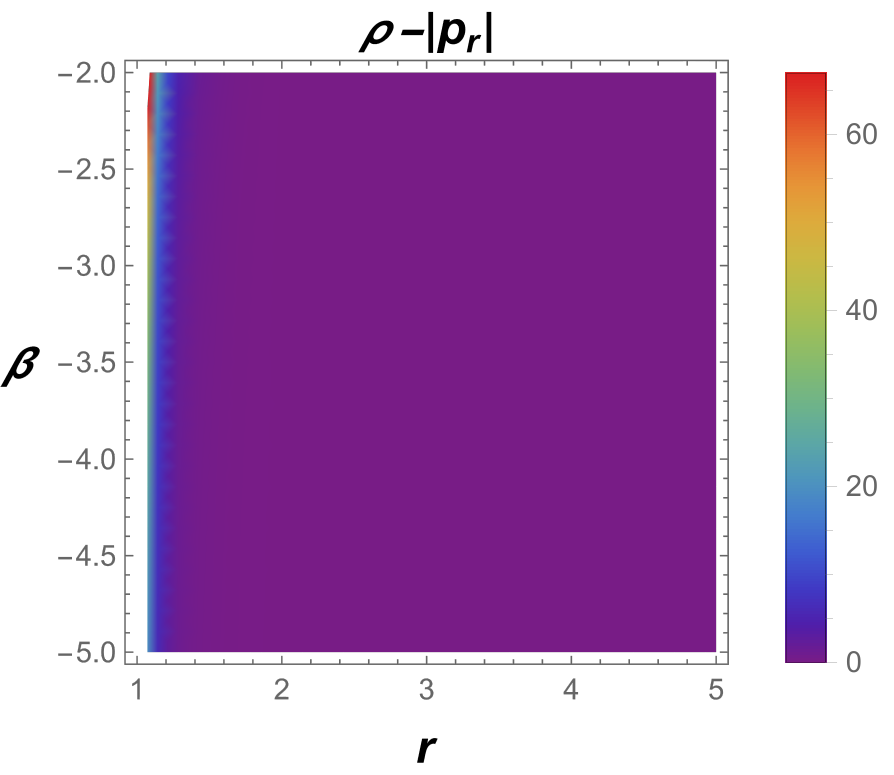}}
       \subfloat[DEC $\rho -| p_t|$ \label{fig:Dec4}] {\includegraphics[width=0.301\linewidth]{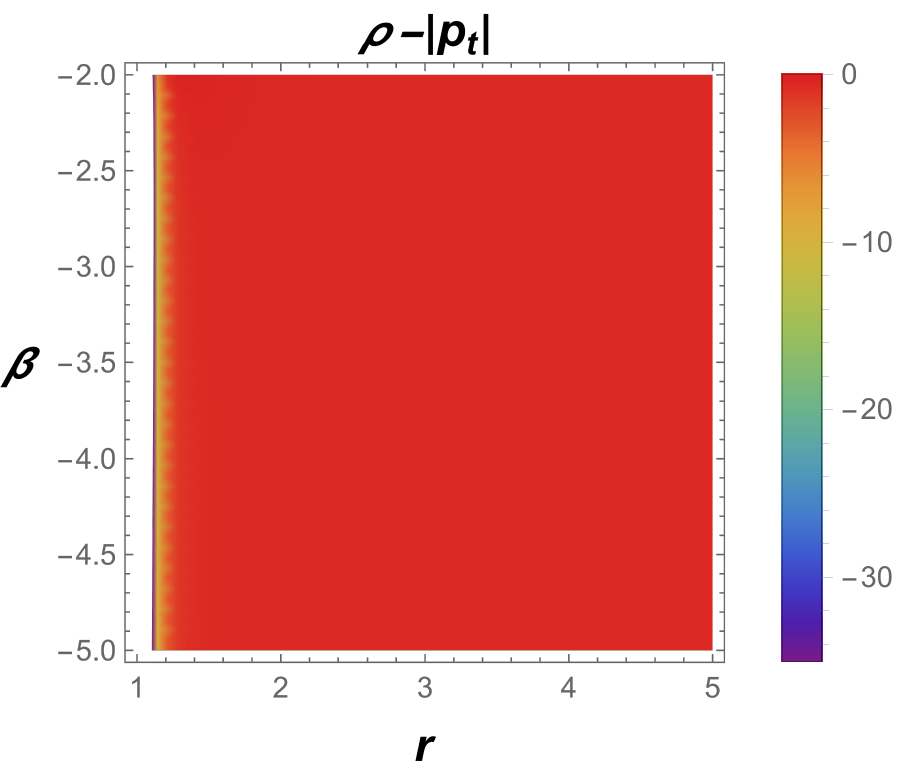}}
        \subfloat[SEC $\rho + p_r + 2p_t$ \label{fig:Dec5}] {\includegraphics[width=0.3\linewidth]{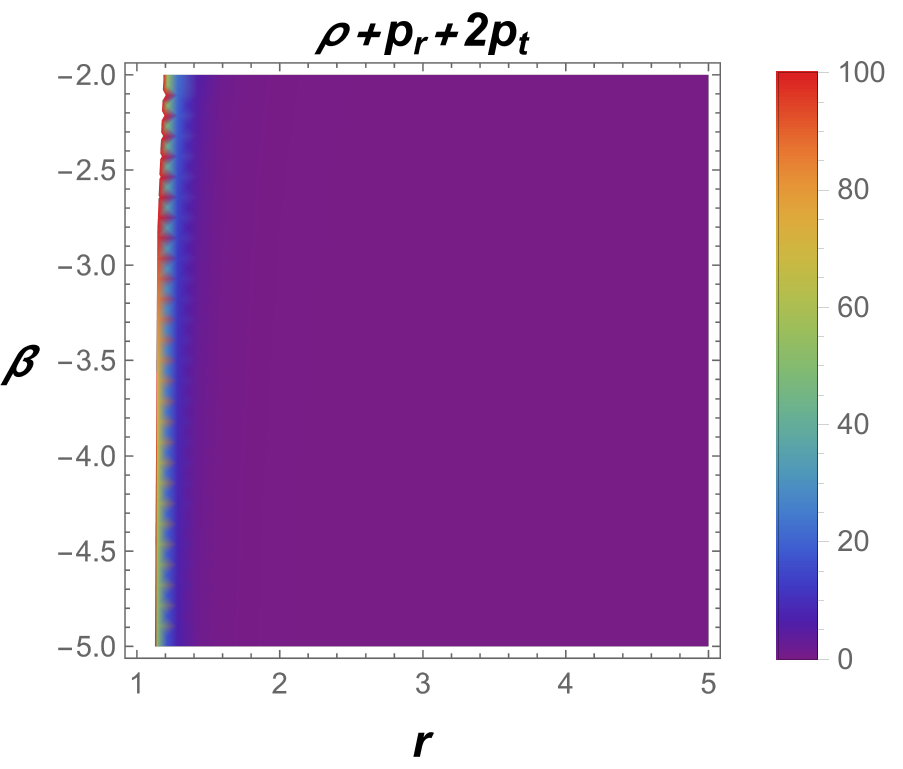}}\\
      \caption{The plots of energy density and ECs with $\alpha=2.5$ and $r_0=1$.}
      \label{fig:nec}
      \end{figure*}
 
\begin{table*}[b!]
		\caption{The overview of energy conditions }
		    \label{tab:table4}
		    \centering
		    \begin{tabular}{|c| c| c| c| c| c| c|}
      \hline
		    $\alpha$ 
		    &\multicolumn{2}{c|}{$\alpha<0$}&
		    \multicolumn{2}{c|}{$\alpha=0$}&
		      \multicolumn{2}{c|}{$\alpha>0$}\\
		      \hline
				 $\beta$ & $(-1, \frac{1}{2})$ & $(3, \infty)$ & $(-\infty, -1)$ & $(\frac{1}{2}, 1\frac{1}{2})$ &  $(-\infty, -1)$ & $(\frac{1}{2}, 1\frac{1}{2})$\\
				\hline
				$\rho$ & Obeyed & Obeyed & Obeyed & Obeyed & Obeyed & Obeyed \\
				$\rho+p_r$ & Violated & Violated & Obeyed & Violated & Obeyed & Violated\\
				$\rho+p_t$ & Obeyed & Obeyed & Violated& Violated & Obeyed & Violated \\
				$\rho-\mid p_r\mid$ & Violated & Violated & Violated & Violated & Obeyed & Violated\\
				$\rho-\mid p_t\mid$ & Violated & Violated & Violated & Violated & Violated & Violated\\
				$\rho+p_r+2 p_t$ & Violated & Obeyed & Obeyed & Violated & Obeyed & Violated\\
    \hline
			\end{tabular}
		\end{table*}
 \end{widetext}
	From Fig.(\ref{fig:nec}), which shows the characteristics behavior of energy density and ECs. The energy density is positive throughout space-time. However, in the non-linear model, NEC is satisfied which implies the avoidance of the necessity of this hypothetical fluid. Also, DEC is violated and SEC is satisfied in Fig.(\ref{fig:nec}). The characteristic of ECs for this case is depicted in Table \ref{tab:table4}.

\section{Geometric embedding of wormhole} \label{section VII}
     The geometric embeddings delineate the visualized insights of WH. Notably, this depends on the choice of the shape function $b(r)$. For spherically symmetric metric, we shall focus on the equatorial slice $\theta=\frac{\pi}{2}$ and the value of the time coordinate is fixed i.e., $t=constant$. On applying these conditions to \eqref{whmetric}, we get
     
     \begin{align}\label{em}
         ds^2 =\dfrac{dr^2}{1-\dfrac{b(r)}{r}  }  + r^2\, d\phi^2.
     \end{align}
    
    Now, one can embed the above slice into its hypersurface with cylindrical coordinates $r, \phi, z$ as
    
    \begin{align}\label{edc}
        ds^2 = dz^2 + dr^2 +r^2 \,d\phi^2.
    \end{align}
    
   Comparing Eqs. \eqref{em} and \eqref{edc}, we can find $z(r)$ that reads
   
   \begin{align}
       \frac{dz}{dr}= \pm \left[ \dfrac{b(r)}{r-b(r)}\right]^{1/2}.
   \end{align}

\begin{figure}[h] 
	    \centering
	    \includegraphics[width=0.7\linewidth]{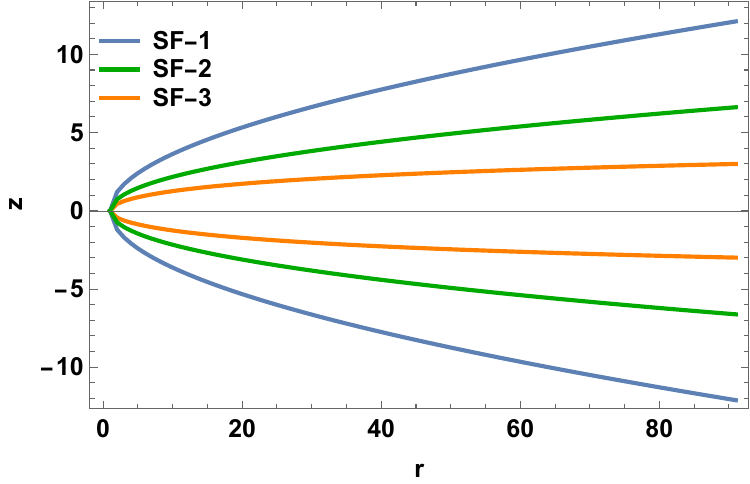}
	    \caption{Two-dimensional embedding diagram.}
	    \label{fig:2DED}
	\end{figure}

\begin{figure}[t!]
      \subfloat[SF-1\label{E1}]
      {\includegraphics[width=0.5\linewidth]{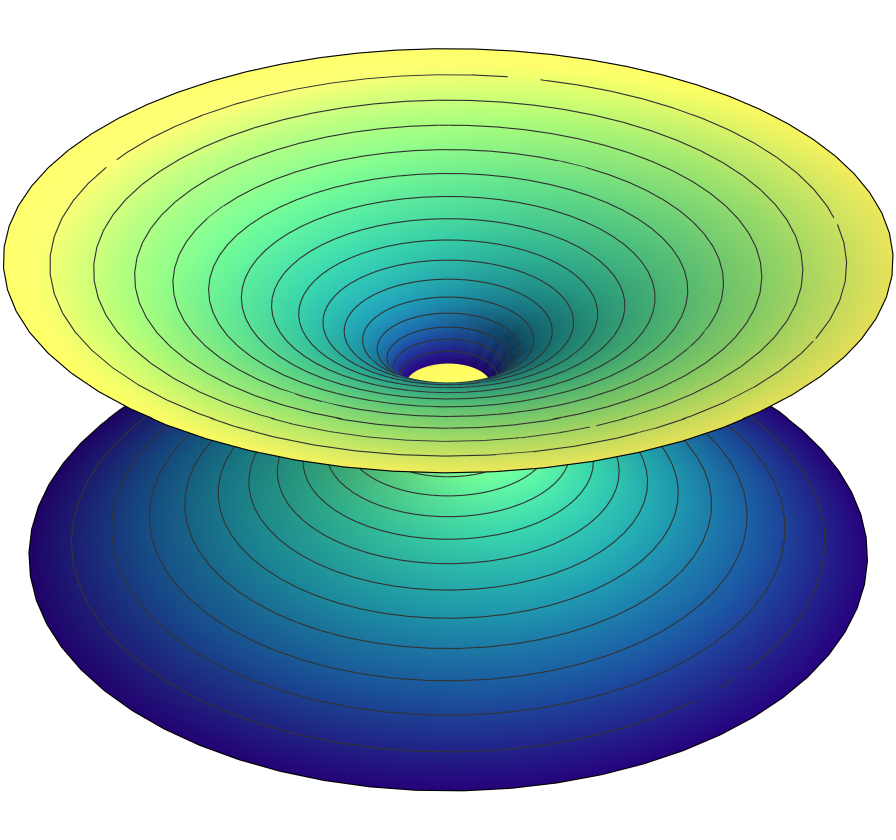}} \\
      \subfloat[SF-2\label{E2}] {\includegraphics[width=0.5\linewidth]{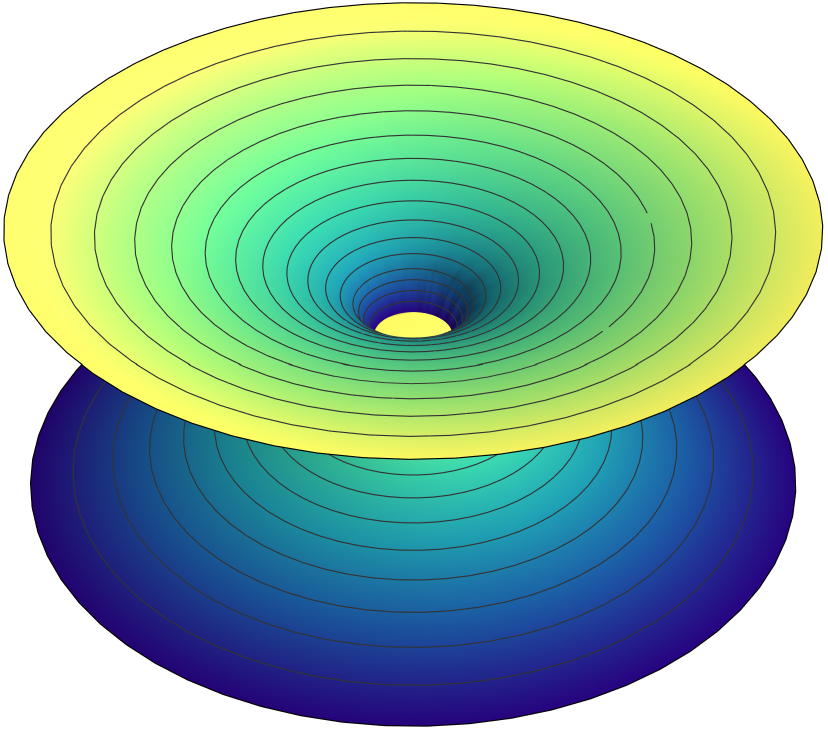}} \\
      \subfloat[SF-3\label{E3}] {\includegraphics[width=0.52\linewidth]{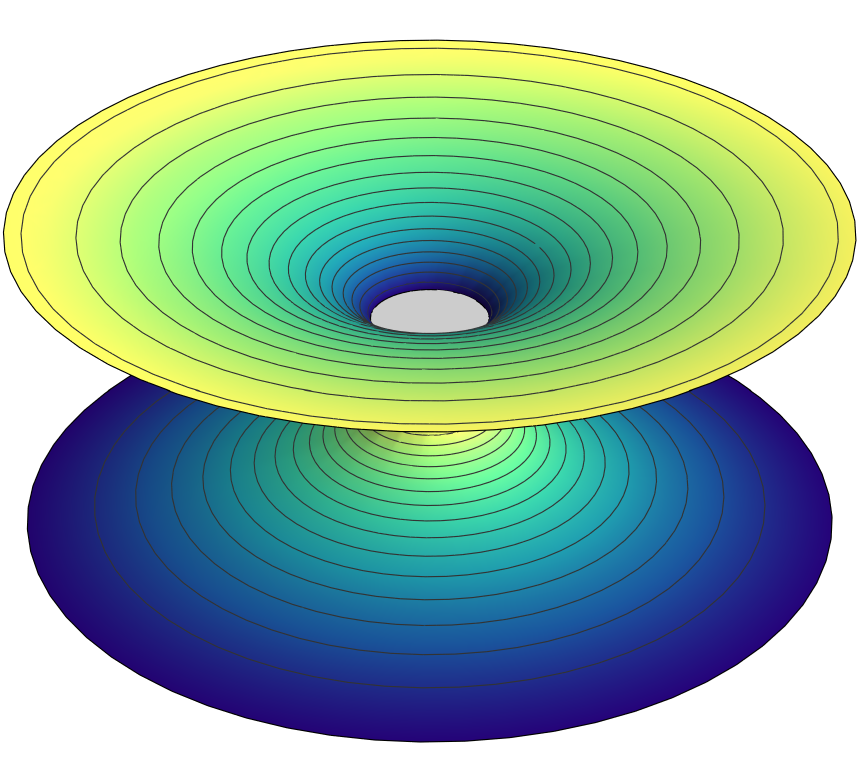}}\\
      \caption{Three-dimensional embedding diagrams. Here SF-1:$b(r) = r_0 \,\dfrac{\log(r + 1)}{\log(r_0 + 1)}$,  SF-2:$b(r)=k+\frac{3 \left(8 \xi ^2+2 \xi -1\right) \rho_0 r^3 \left(\frac{r_0}{r}\right)^n}{(3-n) (4 \xi -3)}$ and SF-3:$b(r) = \frac{r_0^2}{r}$ with $r_0=1, n=5, \xi=-0.4, \rho_0=0.8$ }
      \label{fig:embedding}
      \end{figure}

    Fig.(\ref{fig:2DED} and \ref{fig:embedding}) shows the two-dimensional and three-dimensional embedding diagrams for different WH models of SF-1, SF-2, and SF-3 with values $r_0=1, n=5, \xi=-0.4,\, \rho_0=0.8$. 
   
\section{Discussion and Final Remarks}\label{sectionVIII}
	In recent years, WHs are one of the most fascinating topics in modern cosmology. Beginning with Flamm's solution to the present humanly traversable WH, numerous endeavors have been taking place on compact objects. In the framework of GR, WH describes a shortcut distance to connect various points of the universe. To examine these solutions, the violation of NEC plays a crucial role associated with the exotic matter. The usage of exotic matter would be minimized to obtain a realistic model in favor of the WH. 
	\par In the present work, we have investigated the spherically symmetric Morris-Thorne traversable WHs within the background of extended symmetric teleparallel gravity (i.e., $\mathpzc{f}( \mathcal{Q}, \mathcal{T})$ gravity), where gravitational interaction is described by the coupling of non-metricity $\mathcal{Q}$ and trace of energy-momentum tensor $\mathcal{T}$. $\mathpzc{f}( \mathcal{Q}, \mathcal{T})$ gravity is a recently established modified theory, so many investigations are underway to explore the current interests of  cosmological scenarios. Moreover, $\mathpzc{f}( \mathcal{Q}, \mathcal{T})$ gravity can provide useful insights for the description of the early and late phases of cosmological evolution \cite{arora} and matter-antimatter symmetry \cite{bhatta}. Here, we consider specific sets of shape functions and redshift functions.  The specific shape functions satisfy all the criteria for traversable WH such as throat condition, flaring-out condition and asymptotic flatness condition. The WH to be traversable, one must demand that there are no horizons present, which are identified as the surfaces with $e^{2\Phi}\to 0$, so that $\Phi(r)$ must be finite everywhere. Moreover, the matter distribution is assumed to be anisotropic. The investigation of WH is mainly done in two scenarios using a linear model of $\mathpzc{f}( \mathcal{Q}, \mathcal{T})$ gravity, $\mathpzc{f}( \mathcal{Q}, \mathcal{T}) = \mathcal{Q} + 2 \, \xi \, \mathcal{T}$ and a non-linear model of $\mathpzc{f}( \mathcal{Q}, \mathcal{T})$ gravity, $\mathpzc{f}( \mathcal{Q}, \mathcal{T}) = \mathcal{Q} + \alpha \,\mathcal{Q}^2 + \beta \,\mathcal{T}$.
	\par In a linear model, we studied two cases. At first, using SF-1: $b(r) = r_0 \,\dfrac{\log(r + 1)}{\log(r_0 + 1)}$, we analyze the three sub-cases of specific redshift functions i.e., $ \Phi(r) = \log {\left(1- \dfrac{1}{r} \right)}, \Phi(r)= c$ and $\Phi(r)=\frac{1}{r}$. This form of shape function is considered by Godani in \cite{godani}. In the first case, we found that NEC is satisfied for radial and tangential pressures, which means that no hypothetical fluid is present, while DEC and SEC are violated. The remaining two cases are violations of NEC and DEC. Next, we solve SED: $\rho = \rho_0 \left(\dfrac{r_0}{r}\right)^n$ to get SF-2: $b(r)=k+\frac{3 \left(8 \xi ^2+2 \xi -1\right) \rho_0 r^3 \left(\frac{r_0}{r}\right)^n}{(3-n) (4 \xi -3)}$. This form of energy density used in \cite{swkim,godani,new}. Also, this density is compared with the literature \cite{capozzi}, some new attempts may be carried out. Here, NEC and DEC are violated and SEC is satisfied. Thus, the violation of NEC defines the possibilities of the presence of exotic matter at the WHs throat. The characteristic of ECs was depicted in Table \ref{tab:table1} and \ref{tab:table2}.
	\par In a non-linear model, we considered SF-3: $b(r) = \frac{r_0^2}{r}$ and constant redshift function. SF-3 was proposed in the original work of Lobo et al. \cite{oliv} and this function satisfies all necessary conditions for traversable WH. Significantly, this redshift function becomes a tideless WH. Further, the model parameters $\alpha$ and $\beta$ were also chosen depending on energy density. For instance, the energy density in the range $(-\infty, -1)$ of $\beta$  is positive when $\alpha>0$. Table \ref{tab:table4} represents the tabulated result for ECs. Besides, we found that NEC  and SEC are satisfied whereas DEC is violated. In addition, we have discussed the geometric embedding of WH in Figs.(\ref{fig:2DED} and \ref{fig:embedding}).
    \par In the first gravity model which resembles GR, NEC is satisfied in the first case and in the next three cases, the NEC is violated, which denotes the presence of a hypothetical fluid. Although, in a non-linear model, NEC is satisfied (\cite{oliv,bane,habib}  for similar results) which denotes the absence of a hypothetical fluid. These are the main advantages of this gravity model.
	\par To conclude, we investigated traversable WH in $\mathpzc{f}( \mathcal{Q}, \mathcal{T})$ gravity, and interestingly our model upholds the presence of exotic matter at the WH throat. In the future, we can examine some more observational interpretations of the WH solution with $\mathpzc{f}( \mathcal{Q}, \mathcal{T})$ gravity models.

\begin{acknowledgments}
  V.V, C.C.C, and N.S.K acknowledge DST, New Delhi, India, for their financial support for research facilities under DST-FIST-2019. Authors are very much grateful to the honorable referee for the illuminating suggestions that have significantly improved our work in terms of research quality, and presentation.
\end{acknowledgments}

	
\end{document}